\documentclass[twocolumn,trackchanges]{aastex701}
\shorttitle{WRLOF model on BH+Be star}
\shortauthors{Z. Li et al.}

\begin{document}

\title{Formation of Be stars via wind accretion: Case study on Black hole + Be star binaries}

\correspondingauthor{Z. Li; X. Chen}

\author[0000-0002-1421-4427]{Zhenwei Li}
\affiliation{International Centre of Supernovae (ICESUN), Yunnan  Key Laboratory of Supernova Research, Yunnan Observatories, Chinese Academy of Sciences (CAS), Kunming 650216, People's Republic of China}
\affiliation{University of Chinese Academy of Sciences, Beijing 100049, People's Republic of China}
\email[show]{lizw@ynao.ac.cn}

\author{Shi Jia}
\affiliation{International Centre of Supernovae (ICESUN), Yunnan  Key Laboratory of Supernova Research, Yunnan Observatories, Chinese Academy of Sciences (CAS), Kunming 650216, People's Republic of China}
\affiliation{University of Chinese Academy of Sciences, Beijing 100049, People's Republic of China}
\email{jiashi@ynao.ac.cn}

\author[0000-0003-0821-4583]{Dandan Wei}
\affiliation{Institute of Science and Technology Austria: Klosterneuburg, Lower Austria, AT}
\email{dandan.wei@hotmail.com}

\author[0000-0002-1421-4427]{Hongwei Ge}
\affiliation{International Centre of Supernovae (ICESUN), Yunnan  Key Laboratory of Supernova Research, Yunnan Observatories, Chinese Academy of Sciences (CAS), Kunming 650216, People's Republic of China}
\affiliation{University of Chinese Academy of Sciences, Beijing 100049, People's Republic of China}
\email{gehw@ynao.ac.cn}

\author{Hailiang Chen}
\affiliation{International Centre of Supernovae (ICESUN), Yunnan  Key Laboratory of Supernova Research, Yunnan Observatories, Chinese Academy of Sciences (CAS), Kunming 650216, People's Republic of China}
\affiliation{University of Chinese Academy of Sciences, Beijing 100049, People's Republic of China}
\email{chenhl@ynao.ac.cn}

\author{Yangyang Zhang}
\affiliation{Zhoukou Normal University, East Wenchang Street, Chuanhui District, Zhoukou, 466001, People's Republic of China}
\email{zhangyy@zknu.edu.cn}

\author[0000-0001-5284-8001]{Xuefei Chen}
\affiliation{International Centre of Supernovae (ICESUN), Yunnan  Key Laboratory of Supernova Research, Yunnan Observatories, Chinese Academy of Sciences (CAS), Kunming 650216, People's Republic of China}
\affiliation{University of Chinese Academy of Sciences, Beijing 100049, People's Republic of China}
\email[show]{cxf@ynao.ac.cn}

\author[0000-0001-9204-7778]{Zhanwen Han}
\affiliation{International Centre of Supernovae (ICESUN), Yunnan  Key Laboratory of Supernova Research, Yunnan Observatories, Chinese Academy of Sciences (CAS), Kunming 650216, People's Republic of China}
\affiliation{University of Chinese Academy of Sciences, Beijing 100049, People's Republic of China}
\email{zhanwenhan@ynao.ac.cn}


\begin{abstract}
Be stars are rapidly rotating main-sequence (MS) stars that play a crucial role in understanding stellar evolution and binary interactions. In this letter, we propose a new formation scenario for black hole (BH) + Be star binaries (hereafter BHBe binaries), where the Be star is produced through the Wind Roche Lobe Overflow (WRLOF) mechanism. Our analysis is based on numerical simulations of the WRLOF process in massive binaries, building upon recent theoretical work.
We demonstrate that the WRLOF model can efficiently form BHBe binaries under reasonable assumptions on stellar wind velocities. Using rapid binary population synthesis, we estimate the population of such systems in the Milky Way, predicting approximately $\sim$ {1800-3200} currently existing BHBe binaries originating from the WRLOF channel. These systems are characterized by high eccentricities and exceptionally wide orbits, with typical orbital periods exceeding 1000 days and a peak distribution around $\sim$10000 days. Due to their long orbital separations, these BHBe binaries are promising targets for future detection via astrometric {and interferometric} observations.

\end{abstract}




\section{Introduction}
\label{sec:1}

Be stars represent the most rapidly rotating class of non-degenerate stars, exhibiting characteristic $H_{\alpha}$ emission in their spectra (\citealt{Quirrenbach1997, Meilland2007, Rivinius2013, Vioque2018, Rivinius2024a}). Binary interaction is widely regarded as the dominant mechanism for explaining the rapid rotation of most Be stars \citep[e.g.,][]{Kriz1975,Packet1981,vanBever1997,Pols1991,McSwain2005}. In the classical binary evolution framework, the accretor gains spin angular momentum through mass transfer from its companion via stable Roche lobe (RL) overflow (hereafter the RL channel). This process has been observationally and theoretically validated as an efficient mechanism for producing Be star + stripped star binaries, and Be X-ray binaries \citep[e.g.,][]{Shaoy2014,Grudzinska2015,Vinciguerra2020,Wangl2018,Shaoy2020,Shaoy2021,Wangl2023,Liub2024,Rocha2024,Bao2025}.

In comparison to mass transfer through RL overflow, wind interaction represents another crucial mode of interaction in binary systems \citep[e.g.][]{Mohamed2012,Abate2013,Skopal2015,Zhekov2021,Kashi2022,Sunm2024}. Whether wind interaction can effectively spin up the accretor and lead to the formation of a Be star remains an intriguing yet relatively underexplored question (e.g., \citealt{Kashi2023}). In this Letter, we investigate the role of Wind Roche Lobe Overflow (WRLOF) in the formation of Be stars. Originally introduced by \citet{Mohamed2007} to explain observed mass outflows in Mira binaries (\citealt{Mohamed2010, Mohamed2012}), the WRLOF model describes a scenario in which the primary star evolves within its RL while its stellar winds become gravitationally focused toward the secondary star (see \citealt{chenx2024} for a recent review). This process is significantly more efficient than the classical Bondi-Hoyle-Lyttleton (BHL) wind accretion mechanism (\citealt{Hoyle1939, Bondi1944}).

Subsequent studies have successfully applied the WRLOF framework to a wide range of binary systems containing asymptotic giant branch (AGB) or red giant branch (RGB) stars (\citealt{Abate2013, Skopal2015, Shagatova2016, Shagatova2021, Skopal2023, Sunm2024}), where the slow, dense winds of evolved stars naturally facilitate WRLOF conditions. Hydrodynamic simulations have further refined the model, quantifying the wind accretion efficiency as a function of wind velocity and binary mass ratio \citep[e.g.,][]{Liuz2017,Chenz2017,Saladino2018,Saladino2019}. Notably, \citet{ElMellah2019b} extended the WRLOF model to massive star binaries, providing a plausible explanation for observed high-mass X-ray binaries, including ultraluminous X-ray sources (\citealt{Zuoz2021, Wiktorowicz2021}).

In this letter we explore the formation of Be stars through the WRLOF model, with a focus on the particularly rare class of black hole (BH) + Be star binaries (hereafter BHBe binaries). Currently, only three candidate systems have been observationally identified-MWC 656 (\citealt{Casares2014}), AS 386\footnote{{AS 386 is not a classical Be star, but rather belongs to the FS CMa type of objects characterized by circumstellar dust. The formation mechanism of such systems remains poorly constrained. See \citet{Schneider2025} for a recent review.}} \citep{Khokhlov2018} and ALS 8814 (\citealt{An2025})—though the nature of ALS 8814 and MWC 656 remains debated (\citealt{Rivinius2024b,Janssens2023,Elbadry2025}). Given the scientific importance of these rare systems, we propose a novel formation scenario for BHBe binaries and present theoretical predictions to guide future observations.

This study is organized as follows. In Section~\ref{sec:2}, we describe the simulated methods including the details of WRLOF model and important input parameters. The results of detailed binary evolutionary simulations and the rapid binary population synthesis model are provided in Section~\ref{sec:3} {and Section~\ref{sec:4}}. Finally, In Section~\ref{sec:5}, we discuss our results and give some conclusions.

\section{Methods and Inputs}
\label{sec:2}

In this study, we utilize the Modules for Experiments in Stellar Astrophysics (MESA, version 12115; \citealt{Paxton2011,Paxton2013,Paxton2015,Paxton2018,Paxton2019}) to model binary star evolution and examine Be star formation through the WRLOF mechanism. Our approach employs MESA's binary module to simultaneously evolve both stellar components within each system. The main inputs are introduced as follows.

\subsection{{Stellar-model physics}}
\label{sec:2.1}

We consider the solar metallicity of $Z=0.0142$ and helium mass fraction of $Y=0.2703$ \citep{Asplund2009}. The OPAL type II opacity table is adopted \citep{Iglesias1996}. We use the mixing length theory with a mixing length parameter of $\alpha_{\rm mlt}=1.8$ \citep{Henyey1965}. Semi-convection is modelled with an efficiency factor of $\alpha_{\rm sc}=0.1$ \citep{Choi2016}. The overshooting is implemented as a step function up to 0.335 times the pressure scale height, as calibrated by \citet{Brott2011}. The implementation of stellar winds refers to \citet{Yoon2006}, where the mass-loss rates for hydrogen-rich stars (surface helium abundance $Y_{\rm s}<0.4$) is computed as in \citet{Vink2001}, while for hydrogen-poor stars ($Y_{\rm s}>0.7$) we used the recipe of \citet{Hamann1995} multiplied by a factor of one tenth. In the range of $0.4<Y_{\rm s}<0.7$, the mass loss rate was interpolated between the two. 

\subsection{{Implementation of the WRLOF Model}}
\label{sec:2.2}
{Although the primary star does not fill its {RL}, the wind can be significantly beamed and bent by the orbital effects. With radiative cooling, the accreted flow can circularize with a low wind speed and then form a disk-like structure \citep{Ducci2009,ElMellah2017,ElMellah2018,ElMellah2019a,ElMellah2019b}. The wind-captured disk enables the accretor to accumulate material and be spun-up. One crucial question is the specific values of the wind accretion efficiencies during the wind accretion process. In this letter, we model the wind accretion process following the numerical simulations of \citet{ElMellah2019b}. The details of their models are introduced as follows.}

{\citet{ElMellah2019b} determined the wind accretion efficiency by modelling the ballistic trajectories of test masses within the co-rotating frame of a binary system. Their approach integrates the equation of motion, accounting for both gravitation forces, non-inertial forces, and wind acceleration, for particles originating from the donor star's surface (see also \citealt{ElMellah2017}). Crucially, they iteratively refine the initial launch positions to accurately trace these trajectories. Accretion is deemed to occur when a trajectory enters the accretor's extended accretion sphere, thereby defining the final wind accretion efficiency. In this letter, we incorporate the wind accretion efficiencies from \citet{ElMellah2019b} into the MESA code. This efficiency depends primarily on four key parameters:}

\begin{itemize}
  \item Speed ratio: $\eta_{\rm wind} \equiv v_{\rm wind}/v_{\rm orb}$, where $v_{\rm wind}$ is the terminal wind speed of the primary and $v_{\rm orb}$ is the orbital speed. A higher $\eta_{\rm wind}$ value would lead to a lower wind accretion efficiency, particularly, the wind accretion is nearly neglected for $\eta_{\rm wind}\gtrsim 20$.  
  \item {RL} filling factor: $f_{\rm RL}\equiv R_{\rm 1}/R_{\rm RL,1}$, where $R_{\rm 1}$ is the primary's radius and $R_{\rm RL,1}$ is its RL radius. A lower $f_{\rm RL}$ value would lead to a lower wind accretion efficiency. 
  \item Mass ratio: $q\equiv M_1/M_2$, where $M_{\rm 1}$ and $M_{\rm 2}$ are the primary and secondary mass. A higher $q$ value would lead to a lower wind accretion efficiency ({For a given primary star, a more massive secondary results in a larger wind capture region, which generally enhances the wind accretion efficiency \citep{ElMellah2019b}.} 
    \item Radial-velocity profile: Determines the wind acceleration and significantly affects accretion efficiency. In general, the profile can be approximated by a $\beta-$law \citep{Puls2008}: 
\begin{eqnarray}
  v_{\beta}(r)=v_{\rm wind}(1-R_{1}/r)^{\beta}.
  \label{eq:1}
\end{eqnarray}
The exponent $\beta$ is rather uncertain due to the rare observations \citep[e.g.,][]{Gimenez2016,Sander2017}. \citet{ElMellah2019b} simulated two cases of $\beta$ values, i.e., 1 and 2, here we adopt the case of $\beta=2$. A smaller $\beta$ value would lead to a lower wind accretion efficiency (see \citealt{ElMellah2019b} and \citealt{Zuoz2021} for more details). 
\end{itemize}

For a fixed $\beta$ value, we determine the wind accretion efficiency by linearly interpolating the data tables provided by \citet{ElMellah2019b}, following the methodology employed by \citet{Zuoz2021} in their study of ultraluminous X-ray sources. The speed ratio $\eta_{\rm wind}$ depends primarily on the terminal wind speed $v_{\rm wind}$ of the primary star, which exhibits substantial variation across stellar types due to fundamentally different driving mechanisms. For example, the winds of Wolf-Rayet (WR) star and OB star are typically driven by metal lines, the wind speed could be as large as several thousand $\rm km\;s^{-1}$. The wind driving mechanism of cool supergiants is usually assumed to be radiation pressure on dust and the wind speed can be as low as several tens $\rm km\;s^{-1}$ \citep{Vink2022}. Following \citet{Hurley2002}, we parameterize the wind speed as a fraction of $\beta_{\rm wind}$ of the surface escape velocity for the mass-losing star (see also \citealt{Belczynski2008}) as 
\begin{eqnarray}
  v_{\rm wind}^2 = 2\beta_{\rm wind}\frac{GM_1}{R_1}.
  \label{eq:2}
\end{eqnarray}
where $G$ represents the gravitational constant.  

\begin{figure}
    \centering
    \includegraphics[width=\columnwidth]{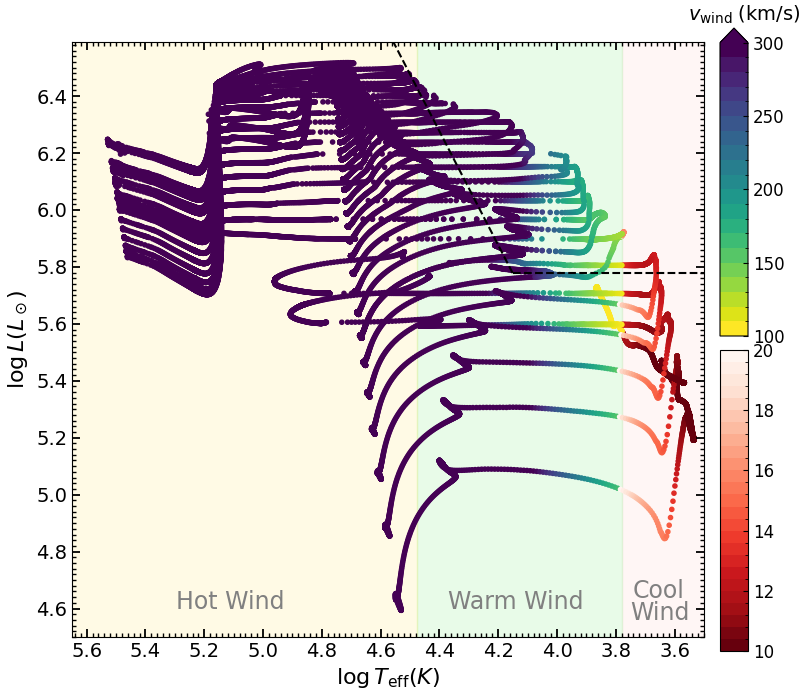}
    \caption{The terminal wind speeds for stars in the fiducial model ($f_{\rm w} = 1$). The initial star masses are from {20} to $150\,M_\odot$ with a step of $5\,M_\odot$ from bottom to the top. The stellar winds according to the effective temperature of the star are divided into three regions: hot wind $T_{\rm eff}>30000\,\rm K$, warm wind ($6000<T_{\rm eff}<30000\,\rm K$) and cool wind ($T_{\rm eff}<6000\,\rm K$). The color bar indicates the specific values of terminal wind speeds, with the displayed range limited to velocities below $300 \,\rm km/s$. The dashed line is the so-called Humphreys-Davidson limit \citep{Humphreys1979}.}
    \label{fig:1}
\end{figure}

To account for the diverse wind properties across stellar evolution phases, we implement a step function for $\beta_{\rm wind}$ with four cases according to the surface helium mass fraction $Y_{\rm s}$ and the effective temperature $T_{\rm eff}$: WR wind ($Y_{\rm s}\geq0.4$), hot wind ($T_{\rm eff}>30000\;\rm K$), warm wind ($6000<T_{\rm eff}<30000\;\rm K$) and cool wind ($T_{\rm eff}<6000\;\rm K$). For the WR wind and hot wind, the $\beta_{\rm wind}$ values are obtained through mass-dependent linear interpolation following \citet{Zuoz2021}. For warm wind and cool wind, we adopt 
\begin{eqnarray}
  \beta_{\rm wind} = \left\{\begin{array}{lrl}
    f_{\rm w}\cdot 1, \;\,{\rm warm\;\,wind},\\
    f_{\rm w}\cdot 1/8, \;\,{\rm cool\,\; wind},
\end{array}\right.
\label{eq:3}
\end{eqnarray}
where $f_{\rm w}$ is a scaling factor ($f_{\rm w} = 1$ in our fiducial model). {In Figure~\ref{fig:1}, we present the characteristic terminal wind speeds implemented in the fiducial model.} The typical wind speeds for red supergiant (RSG) winds are in the range of $10-20\;\rm km\;s^{-1}$, and for luminous blue variable (LBV) winds are in the range of $100-300\;\rm km\;s^{-1}$. Although these values may represent underestimates, they remain consistent with observations within the same order of magnitude (e.g., see Table $1$ in \citealt{Vink2022}). We explore enhanced wind speeds with $f_{\rm w}=1.5,\;2$ and $3$ in Section~\ref{sec:3.2}.

In the numerical simulations, we found the adopted wind speeds for hot winds, WR winds and warm winds in stars with initial masses {of $20-40\,M_\odot$} have negligible effects on the wind accretion processes. Instead, efficient wind accretion predominantly occurs in two distinct cases: (1) the cool winds of RSG in the {$\sim 20-40\,M_\odot$} range and (2) the warm winds of LBV in more massive stars. This behavior arises because wind accretion efficiency exhibits a strong positive correlation with the RL filling factor ($f_{\rm RL}$). Given that the initial binary separations are sufficiently wide to avoid RL overflow, efficient accretion is primarily triggered when the primary star's radius approaches its RL radius (typically $f_{\rm RL}\gtrsim 0.4$; see Figure~\ref{fig:2} below). Consequently, only the terminal wind speeds of stars near their maximum radial expansion, such as RSGs in the $\sim 20-40\,M_\odot$ range and LBVs in more massive systems, significantly influence our results. {Our model does not incorporate special treatment for LBV–type winds, whose strength can influence the wind accretion process to some extent. This issue is addressed in detail in the Appendix~\ref{sec:A}.}

\begin{figure*}
  \centering
    \begin{minipage}[t]{0.35\textwidth}
        \centering
        \includegraphics[width=\textwidth]{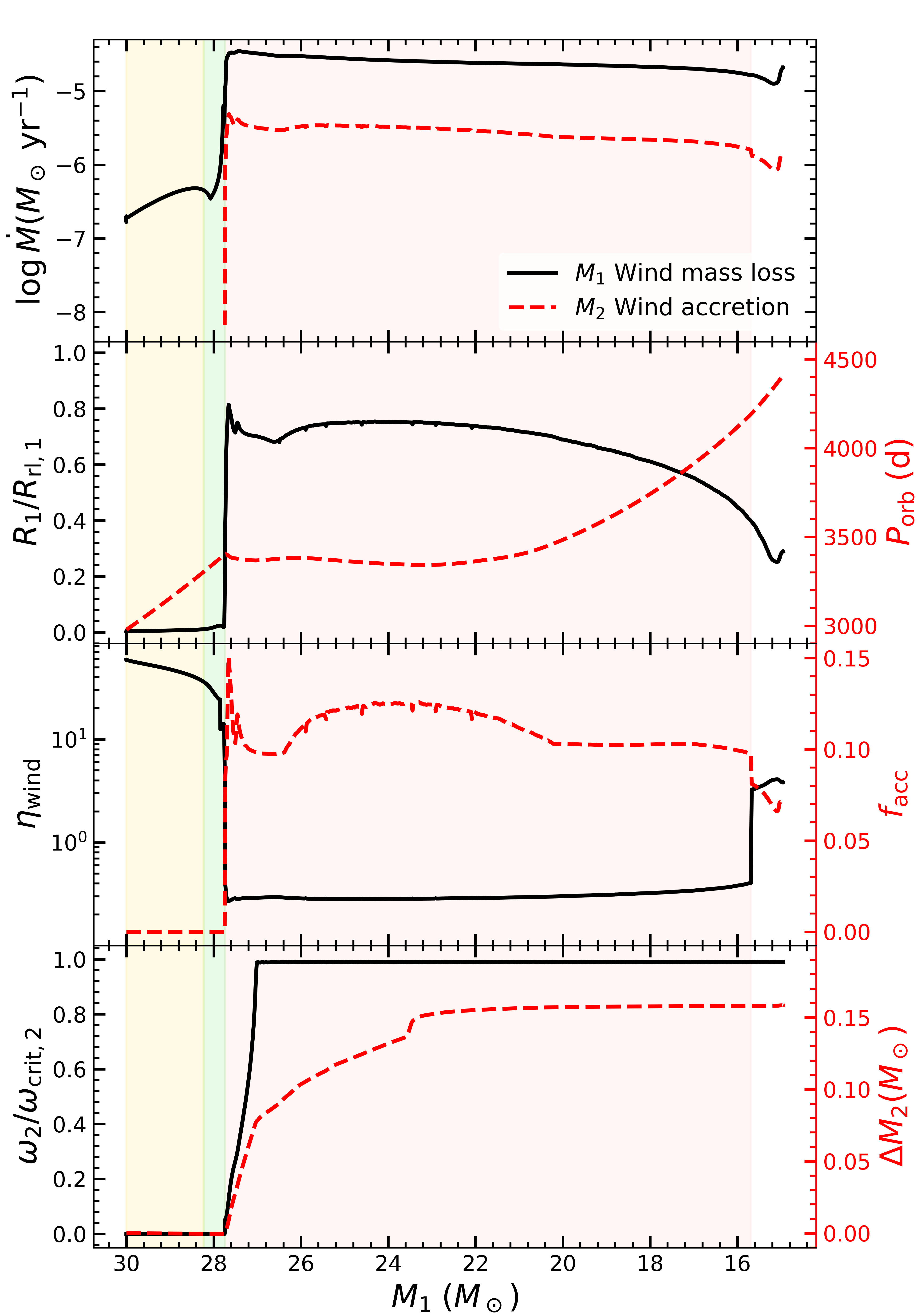}
    \end{minipage}
    \begin{minipage}[t]{0.35\textwidth}
        \centering
        \includegraphics[width=\textwidth]{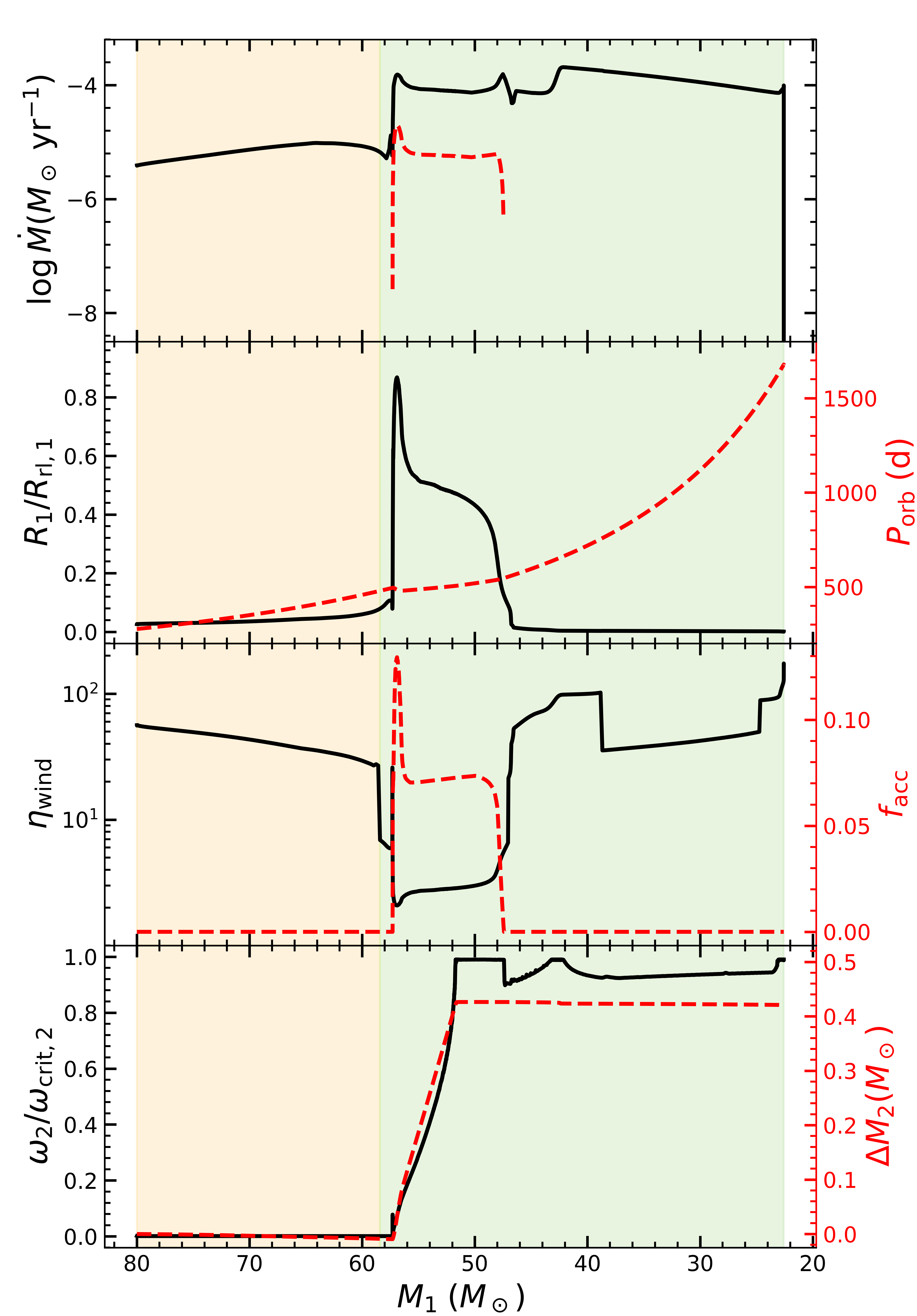}
    \end{minipage}
    \caption{Two evolutionary examples for the formation of BHBe binaries in the WRLOF model. The initial binary parameters are $M_{\rm 1}=30\,M_\odot,\;M_{\rm 2}=5\,M_\odot,\;P_{\rm orb}=2977.6\,\rm d$ for the left column, and $M_{\rm 1}=80\,M_\odot,\;M_{\rm 2}=10\,M_\odot,\;P_{\rm orb}=277\,\rm d$ for the right column. The black solid lines and red dashed lines in the first row represent the wind mass-loss rates of the primary stars and wind accretion rates of the secondary stars, respectively. The second row shows the ratio between primary radius and primary RL radius. The black solid lines and red dashed lines in the third row correspond the speed ratio $\eta_{\rm wind}$ and the wind accretion efficiencies, and in the last row are for $\omega_{\rm 2}/\omega_{\rm crit,2}$ and accumulated masses of secondary stars, respectively. The colored regions represent the same features as those in Figure~\ref{fig:1}.}
    \label{fig:2}
\end{figure*}

\subsection{{Binary grid and BH formation}}
\label{sec:2.3}

We conduct a comprehensive parameter study of binary systems by evolving models with primary (BH progenitor) masses ranging from {20} to $100\,M_\odot$\footnote{{On the basis that stars below 20 $M_\odot$ more frequently yield NSs \citep{Ertl2016,Heger2023,Maltsev2025}, we exclude them; for the higher-mass range ($20-100\,M_\odot$), where NS formation remains possible \citep[e.g.][]{Ertl2016}, we implement a simplification by assuming all remnants are BHs.}} in a step of $10\,M_\odot$, and secondary (Be star progenitor) masses of $5,\;10,\;15$ and $20\,M_\odot$. The initial orbital periods are carefully selected based on the maximum evolutionary radii of the primary stars. {We first identify the minimum orbital period via the bisection method in our MESA simulations, guaranteeing that the primary star will fill its {RL} for any initial period shorter than this threshold.} {The} maximum orbital period is $25$ times this minimum value ({this choice ensures that the entire range of initial orbital periods that produce Be stars is covered}). We simulate 50 binary systems with logarithmically spaced periods between these limits. The initial parameter grid was constructed to efficiently cover most of the BHBe binary system configurations without excessive computational burden. For convenience, {we assume that all binary orbits are initially circular}. 

For the primary stars (BH progenitors), we stop the simulations as the central carbon is exhausted. {In our simulations, the final black hole mass is calculated following the delayed supernova (SN) mechanism, using the fitting formula from \citet{Fryer2012}, in which the BH mass is primarily determined by the pre-SN mass and the corresponding C/O core mass.} {During the BH formation phase, the BH may receive a natal kick, which affects the binary configuration of the BHBe binary. In our simulations, }the BH kicks are scaled by a factor of $1.4\,M_\odot/M_{\rm BH}$ based on the neutron star (NS) kicks $\nu(\rm k)$ \citep{Fragos2023}, where $\nu_{\rm k}$ is drawn from a isotropic Maxwellian distribution with dispersion $\sigma=265\;\rm km\;s^{-1}$ \citep{Hobbs2005}. {The post-kick orbital parameters were determined using the kick formalism of \citet{Hurley2002}.}

\subsection{{Angular momentum accretion and Tides}}
\label{sec:2.4}

We model the angular momentum transfer from accreted wind material assuming a Keplerian disk, where the specific angular momentum is given by $(GM_2R_2)^{1/2}$ \citep{Hurley2002,deMink2013}, with $R_2$ representing the accretor's radius. When the accretor approaches the critical rotation ($v_{\rm rot}/v_{\rm crit} \approx 0.99$; $v_{\rm rot}$ is the rotation velocity and $v_{\rm crit}$ is the critical rotation velocity), we implement enhanced mass loss following \citet{Petrovic2005}. The Be star is born when the rotation velocity exceeds $0.7v_{\rm crit}$ and the lifetime of Be star is assumed to be the {remaining main-sequence (MS)} lifetime\footnote{{Here we employ non-rotating stellar models to calculate the MS lifetime for a given stellar mass. Be stars are rapid rotators, and rotation generally extends their MS lifetimes due to rotational mixing, which refuels the core with fresh hydrogen. As simulated in \citet{Ekstrom2012}, rotation can increase the MS lifetime by up to $\sim 25\%$ compared to non-rotating models. This implies that our predicted numbers of BHBe binaries may be somewhat underestimated.}}of a star of equivalent mass (\citealt{Shaoy2014}).

We incorporate tidal effects following \citet{Fragos2023} but adopt a small tidal damping factor of $F_{\rm tid} = 0.01$. {The small $F_{\rm tid}$ value effectively minimizes tidal interactions in our simulations. A larger tidal damping factor could potentially alter the parameter space for Be stars in the WRLOF model by reducing orbital separations during wind accretion (Section~\ref{sec:3.2}; see also \citealt{Kotko2024} for detailed discussions with a more effective tide). We intentionally minimize tidal interactions to focus on the WRLOF mechanism}, based on the observations for giant binaries of \citet{Niej2017}.\footnote{{All MESA input files are available at \dataset[10.5281/zenodo.17588037]{\doi{10.5281/zenodo.17588037}}.}}

\begin{figure*}[ht!]
    \centering
    \includegraphics[width=0.7\textwidth]{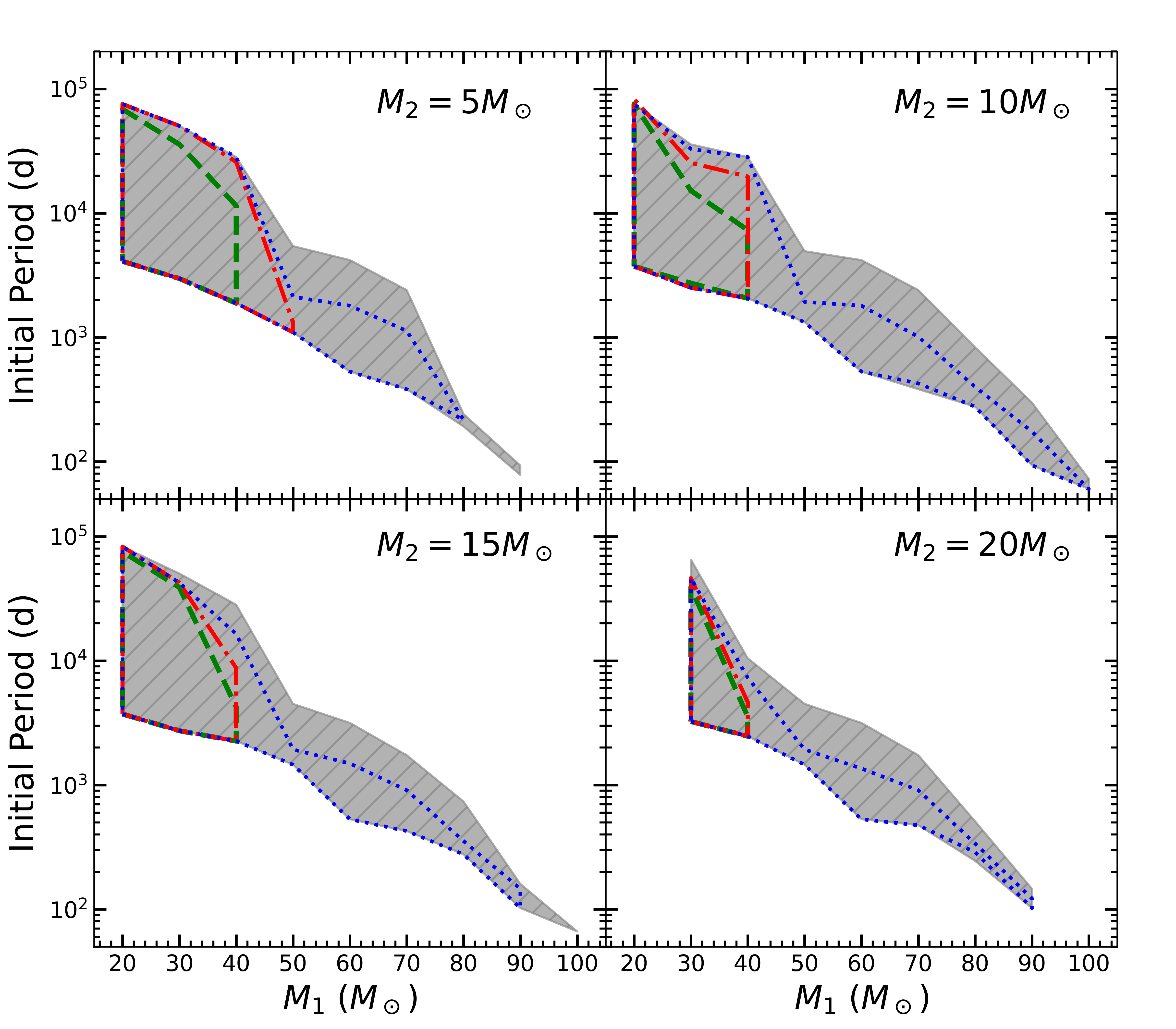}
    \caption{The parameter spaces for BHBe binaries in the WRLOF model. The grey shaded regions represent our fiducial model of $f_{\rm w} = 1$. Other cases with higher $f_{\rm w}$ of $1.5,2$ and $3$ are shown in blue dotted, red dash-dotted and {green} dashed contours, respectively. The primary stars below the lower boundaries would fill their RLs. For binaries above the upper boundaries, the accretion efficiencies are too low to form the Be stars due to the small RL filling factors. }
    \label{fig:3}
\end{figure*}

\section{{Detailed binary evolution simulations}}
\label{sec:3}

\subsection{{Representative Evolutionary Pathways}}
\label{sec:3.1}

In Figure~\ref{fig:2}, we present two characteristic examples of wind accretion processes in BHBe binary formation. The left column illustrates the evolution of a system with an initial primary mass of $30\,M_\odot$, initial secondary mass of $5\,M_\odot$, and an initial orbital period of $2977.6\;\rm d$. During the MS stage and sub-giant phases, the primary loses approximately $2\,M_\odot$ material. However, the high wind speeds (several hundred to thousand $\rm km\;s^{-1}$), combined with relatively low orbital speeds (several tens $\rm km\;s^{-1}$), result in $\eta_{\rm wind}>10$, making wind accretion negligible in these evolutionary stages. The accretion dynamics change dramatically when the primary evolves into the RSG phase. The stellar radius expands significantly while the effective temperature drops below $6000\;\rm K$, causing a substantial decrease in wind speed, as described by Equation~(\ref{eq:3}). This transition reduces $\eta_{\rm wind}$ to $0.3-0.4$, enabling efficient wind accretion with efficiencies reaching $0.15$. Consequently, the secondary rapidly accumulating $\sim 0.15\,M_\odot$ of material, spinning up to the critical rotation velocity and becomes a Be star. {The simulations were run until central carbon exhaustion, at which point the stellar mass was $14.9\,M_\odot$, applying {the delayed SN mechanism} to this value yields a remnant BH mass of $13.4\,M_\odot$.}

The right column of Figure~\ref{fig:2} presents the case of a more massive binary, with initial parameters of $M_1 = 80\,M_\odot,\; M_2 = 10\,M_\odot$ and $P_{\rm orb}=277\;\rm d$. Unlike the previous example, this massive primary experiences significant mass loss ($\sim 20\,M_\odot$) during its MS phase due to the strong wind, preventing its evolution into the RSG phase (see also \citealt{Kruckow2024}). The primary instead enters the LBV phase with considerable radial expansion ($R_{1}>0.4R_{\rm RL,1}$), during which the corresponding $\eta_{\rm wind}$ value is about $2-3$. This configuration enables efficient wind accretion, allowing the secondary star to accumulate approximately $0.4\,M_\odot$ of material, which spins it up to the critical rotation and initial the Be star phase through angular momentum transfer. {The stellar mass at central carbon exhaustion is approximately $22.5\,M_\odot$, resulting in a final BH mass of $20.3\,M_\odot$. In both examples, the orbital periods exhibit a tendency to increase after wind accretion, as illustrated by the second row of Figure~\ref{fig:2}.}

\subsection{Grid spaces for Be stars}
\label{sec:3.2}

Our simulations systematically investigate binary systems with primary masses ranging from {20} to $100\,M_\odot$, and secondaries of $5,\;10,\;15$ and $20\,M_\odot$. As shown in Figure~\ref{fig:3}, the grey-shaded regions are for the parameter space of successful Be star formation, bounded below by systems undergoing RL overflow and above by systems with insufficient accretion efficiency due to small RL filling factors. For increasing primary mass, the viable initial orbital separations decrease because more massive stars develop smaller maximum radii due to the strong wind. Besides, the parameter space tends to be smaller with the increase of secondary mass, since higher secondary masses require more accreted material for spin-up (owing to greater rotational inertia; see Figure~\ref{fig:2}). Notably, systems with $80-90\,M_\odot$ primaries in the upper left panel typically have narrower parameter space than that of other three cases. The reason is that the wind accretion efficiency generally decreases with the increase of mass ratio (primary/secondary; see Section~\ref{sec:2}). Primaries massive than $100\,M_\odot$ fail to produce Be stars altogether, since the wind is so strong that these stars would rapidly evolves to WR stars which possessing extremely high wind speed \citep{Vink2022}. 

\begin{figure*}
    \centering
    \includegraphics[width=0.8\textwidth]{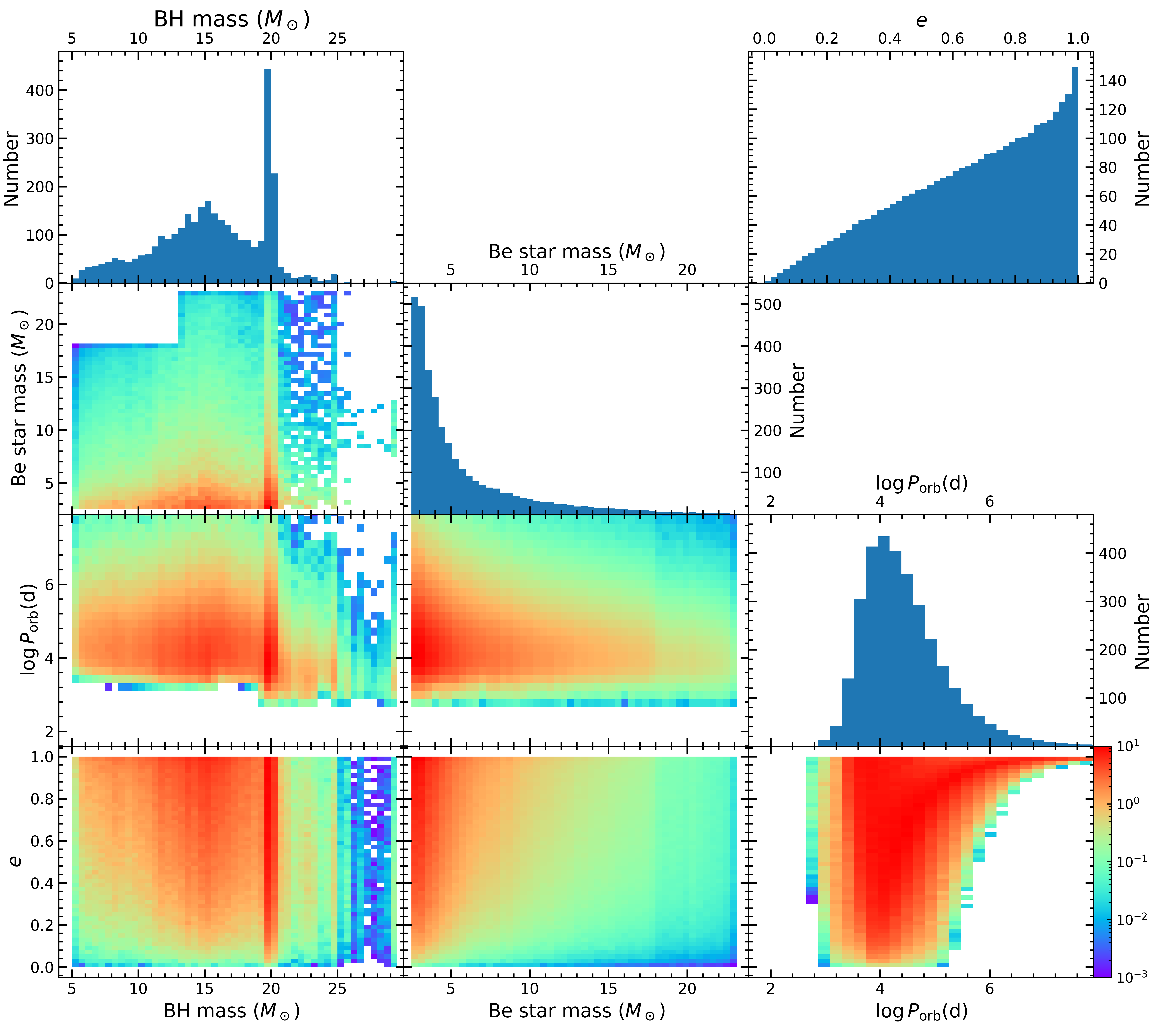}
    \caption{The density distributions of binary parameters for BHBe binaries from WRLOF model. The total number of BHBe binaries in the Galaxy at current epoch is $3225$. 
    Note the color bar is shown in logarithm scale.}
    \label{fig:4}
\end{figure*}

The terminal wind speed of the primary star is the critical parameter governing the parameter spaces of Be stars. We then simulated a series of binaries with enhanced wind speeds, i.e., $f_{\rm w} = 1.5,\;2$ and $3$ for both warm wind and cool wind, as shown in the closed contours of Figure~\ref{fig:3}. We see that the parameter spaces for Be stars decrease with large terminal wind speeds as expected. For primary stars below $40\,M_\odot$, efficient accretion occurs during the RSG phase where the typical $\eta_{\rm wind}(v_{\rm wind}/v_{\rm orb})$ is smaller than $1$ in the fiducial model (as shown in the left column of Figure~\ref{fig:2}). Even with enhanced winds (multiplies a factor $f_{\rm w}$), the terminal wind speeds remain comparable to the orbital speeds ($30-80\;\rm km\;s^{-1}$), allowing sustained accretion. In contrast, more massive systems primarily accrete during the LBV phase, where the typical $\eta_{\rm wind}$ value is about $2-3$ in the fiducial model (see the right column in Figure~\ref{fig:2}). Note that the wind accretion efficiencies drop sharply for $\eta_{\rm wind}\gtrsim 3-4$ \citep{Saladino2019,ElMellah2019b}. A more larger terminal wind speed would result in the significant decrease of wind accretion efficiencies. Therefore, we see that the complete disappearance of viable parameter space for $M_{\rm 1,i}\gtrsim 50\,M_\odot$ when $f_{\rm w}$ exceeds $2$. 

\section{Rapid binary population synthesis}
\label{sec:4}

\subsection{{Inputs and methods}}
\label{sec:4.1}

We perform a comprehensive population synthesis study of BHBe binaries formed through the WRLOF channel in our Galaxy, {restricted to solar metallicity}. We randomly generate $10^9$ primordial binaries based on the following probability distribution function. The initial mass of the primary star follows the initial mass function given by \citet{Miller1979}. A flat distribution of the initial mass ratio (secondary mass/primary mass) is adopted in calculating the secondary mass. The initial binary separation follows the distribution given by \citet{Han1998}. The star formation rate is considered to constant of $5\,M_\odot\,\rm yr^{-1}$ over the past 14 Gyr. 

{Our population synthesis model is built upon the pre-calculated grids shown in Figure~\ref{fig:3}, using interpolation and extrapolation techniques. The progenitor mass of the Be star is assumed to lie between $2.5$ and $22.5\,M_\odot$, consistent with the observed properties of Be stars (corresponding to spectral types A0–O9; e.g., \citealt{Shaoy2021}). The modeling procedure is as follows. For each primordial binary system, we first determine its position in the $M_1-P_{\rm orb}$ parameter space by matching the secondary mass $M_2$ to one of four pre-calculated grid cells (as shown in Figure~\ref{fig:3}). These grids, corresponding to initial secondary masses of $5$, $10$, $15$, and $20\,M_\odot$, span the mass intervals $2.5$–$7.5\,M_\odot$, $7.5$–$12.5\,M_\odot$, $12.5$–$17.5\,M_\odot$, and $17.5$–$22.5\,M_\odot$, respectively. A primordial binary is considered capable of forming a BHBe binary only if its initial binary parameters fall within the corresponding grid cells. For such systems, we derive the binary parameters at the pre-core collapse stage by linearly interpolating between the results of adjacent grid models (the Be star mass is computed as the initial secondary mass plus the accreted mass, where the latter is obtained via interpolation). We then assume that the primary star undergoes direct collapse to form a BH, with its mass given {based on the delayed SN mechanism}. Finally, we account for the effect of BH natal kicks on the binary orbit (as described in Section~\ref{sec:2.3}), modifying the orbital period and eccentricity. We recall our assumption of an initially circular orbit; any orbital eccentricity is thus imparted exclusively by the natal kick.}


\subsection{{Galactic BHBe binary population}}
\label{sec:4.2}

Our population synthesis predicts approximately {1862-3225} BHBe binaries currently exist in the Milky Way, {with the number decreasing with the wind enhancement factor $f_{\rm w}$ (3225, 2886, 2389, and 1862 for $f_{\rm w} = 1,\; 1.5,\; 2,$ and $3$, respectively.)} {The corresponding formation rates decrease from $7.6$ to $4.6\times 10^{-5}\,\rm yr^{-1}$ as $f_{\rm w}$ increases from $1$ to $3$, with specific values of $7.6,\; 6.7,\; 5.2, \;4.6 \times 10^{-5}\,\rm yr^{-1}$ for the respective wind factors.} 

{Figure~\ref{fig:4} presents the density distributions of the binary parameters, including BH mass, Be star mass, orbital period and eccentricity, for the fiducial model $f_{\rm w}=1$.} {The results for other $f_{\rm w}$ values are provided in Figure~\ref{fig:B1} of Appendix~\ref{sec:B}.} The BH masses span $5-30\,M_\odot$ with a primary peak at $\sim 20\,M_\odot$ (from $40-70\,M_\odot$ progenitors) and a secondary peak at $\sim 15\,M_\odot$. {The secondary peak originates from the competition between the initial mass function and the BH natal kicks. While the initial mass function favors a higher abundance of lower-mass stars (and thus lower-mass BH progenitors), the assumed inverse mass dependence of the natal kick means that binaries hosting these lower-mass BHs are more easily disrupted.} The primary peak corresponds to a plateau in BH mass ($\sim 20\,M_\odot$) for progenitor stars with initial masses of $40-70\,M_\odot$ (see Figure~\ref{fig:C1} in Appendix~\ref{sec:C}). This mass plateau results from the competing effects of stellar winds and core burning rates (e.g., \citealt{Choi2016,Fragos2023,Andrews2024,Kruckow2024}). However, we emphasize that significant uncertainties remain in final remnant predictions. Various factors can substantially influence the pre-SN structure, such as convective overshoot, element diffusion and even the numerical parameters (e.g., mesh number and time step). As demonstrated in Figure~\ref{fig:C1}, different MESA simulations produce markedly different relationships between zero-age MS star mass and final BH mass. {Furthermore, our model assumes that the BHs form via direct collapse. Alternative formation scenarios, such as those involving fallback or SN explosions \citep[e.g.,][]{Ertl2016,Heger2023,Maltsev2025}, could somewhat alter the resulting BH mass distribution (while other system properties likely remain largely unchanged, as demonstrated for BH + MS binaries by \citealt{Kruckow2024}).} Consequently, while our results provide valuable insights, the BH mass distribution shown in Figure~\ref{fig:4} requires further verification through additional studies and observational constraints. 

The Be star mass distribution exhibits a characteristic peak at $\sim 3\,M_\odot$ with a monotonically decreasing trend toward higher masses, a direct consequence of the longer MS lifetimes of lower-mass stars (see also \citealt{Kruckow2024}). The orbital period distribution shows a pronounced peak around $10000\,\rm d$, with seldom BHBe binaries having orbital periods shorter than $\sim 1000\,\rm d$. {The long orbital periods originate from the fact that wind interactions tend to widen the orbit, as shown in Figure~\ref{fig:2}.} Furthermore, the eccentricity tends to increase with longer orbital periods. This occurs because in wide-orbit binaries, even a small kick velocity imparted to the BH can result in a highly eccentric orbit (e.g., \citealt{Kruckow2024}). 

\subsection{{Comparison with the classical RL channel}}

{BHBe binaries can also form through the classical RL channel, where the massive primary stably transfers mass to its companion. However, \citet{Shaoy2014} predicted that this channel produces fewer than $\sim 250$ such systems in the Galaxy (similar results were also found in \citealt{Belczynski2009,Grudzinska2015}), approximately an order of magnitude lower than the WRLOF model. The low predicted number stems primarily from the narrow parameter space permitted for stable mass transfer. Specifically, stable RL overflow, necessary to spin up the Be star, requires the system to have a sufficiently low initial mass ratio (primary/secondary). Systems with a high initial mass ratio would undergo unstable mass transfer, which fails to form a rapidly spinning Be star. Consequently, this channel exclusively produces massive Be stars (typically $\gtrsim 10\,M_\odot$; in model II of \citealt{Shaoy2014}, their masses even exceed $\sim 17\,M_\odot$). In contrast, the WRLOF model is not constrained by mass transfer stability. This allows for the formation of numerous BHBe binaries with lower-mass Be stars $\lesssim 10\,M_\odot$. When considering the lifetimes of Be stars, these lower-mass stars contribute more significantly to the overall population. Consequently, the WROF model predicts a much larger number of BHBe binaries than the classical RL channel.}

{The BHBe binaries formed through the WRLOF model exhibit distinct physical characteristics from those produced by the classical RL channel.} In our model, the primary star never fills its RL, therefore, the binary system maintains wide separation throughout the Be star formation. In contrast, If the Be star is produced from RL channel, the BHBe binaries typically have orbital periods in the range of $10-1000\;\rm d$ (\citealt{Belczynski2009,Shaoy2020,Shaoy2021,Rocha2024,Bao2025}; potentially extending to $10^5\;\rm d$) and lower eccentricities ($\lesssim 0.5$; \citealt{Rocha2024}). Thus, the combination of extremely long orbital periods ($\gtrsim 1000\;\rm d$, peaking at $10000\;\rm d$) and higher eccentricities serves as a clear observational signature distinguishing WRLOF-formed BHBe binaries. 

Recent observations by \citet{An2025} have identified a BHBe binary candidate (ALS 8814\footnote{The recent work of \citet{Elbadry2025} suggested that ALS 8814 may not host a BH and could instead be a hierarchical triple system. Nevertheless, since the nature of ALS 8814 does not affect our results, we continue to treat it as a BH + Be star binary.}) with an orbital period of $176.6\;\rm d$ and moderate eccentricity $e=0.23$. This binary's compact orbit lies well outside our predicted parameter space for WRLOF-formed binaries, effectively excluding this formation channel for ALS 8814. The detection of BH binaries in extremely wide orbits presents significant observational challenges, as these systems typically lack X-ray emission \citep[e.g.,][]{Zhangf2004,Raguzova2005,Liub2024,Rocha2024,Sen2024}. However, it is still possible to uncover the wide binaries containing dormant BHs via astrometric {and interferometric} methods. In the recent observations, Gaia DR3 data has verified three wide BH binaries, including Gaia BH1 \citep{Chakrabarti2023,Elbadry2023a}, Gaia BH2 \citep{Elbadry2023b,Tanikawa2023} and Gaia BH3 \citep{Gaia2024}, as well as other unseen compact objects \citep[e.g.,][]{Andrews2022,Wangs2024}. Particularly, Gaia BH3 has a BH mass of $32.7\pm0.82\,M_\odot$, companion star mass of $0.76\pm0.05\,M_\odot$, orbital period of $4253.1\pm98.5\,\rm d$ and eccentricity of $0.7291\pm0.0048$. The orbital parameters of Gaia BH3 align closely with our simulated BHBe binaries in the $P_{\rm orb}-e$ plane (lower right panel in Figure~\ref{fig:4}). This detection demonstrates Gaia's unique capability to identify BH binaries in extremely wide orbits and provides encouraging prospects for uncovering WRLOF-formed BHBe binaries through ongoing astrometric surveys.

\subsection{{Implications for NS/Be binary and single Be star}}

{This study focuses exclusively on Be stars with BH companions. While parameter space likely exists for forming Be stars with NS companions, their production via the WRLOF channel is highly improbable. A NS remnant typically forms when the primary mass is below $\sim 30\,M_\odot$ or potentially even for more massive progenitors under certain conditions \citep{Ertl2016,Heger2023,Maltsev2025}. However, the pre-SN binaries formed in the WRLOF channel are characterized by very wide orbits. Combined with the typically large natal kicks imparted to NSs during formation, these systems have a high probability of being disrupted. To test this, we modeled the outcome for all binary systems in the grids of Figure~\ref{fig:3}, assuming each system produces a NS (fixed at $1.4\,M_\odot$) and a Be star, with NS natal kicks following a Maxwellian distribution ($\sigma=265\;\rm km\,s^{-1}$; \citealt{Hobbs2005}). Applying the population synthesis method from Section~\ref{sec:4.1}, we find that a simulation of $10^9$ primordial binaries produces only {34} surviving NS+Be star binaries. Factoring in the Galactic star formation history and Be star lifetimes, we estimate that fewer than {three} such binary exists in the current Galaxy. Therefore, we conclude that the formation of NS + Be star binaries is highly suppressed in the WRLOF model, and viable production would require natal kicks of extremely low magnitude (e.g., \citealt{Valli2025}). }

{For BHBe binaries formed through the WRLOF channel, We assume that the BH receives a natal kick scaled by a factor of $1.4\,M_\odot/M_{\rm BH}$ relative to the NS kicks. This results in BH kick velocities typically an order of magnitude lower than those of NSs. Nevertheless, even with these reduced kicks, we find that approximately {85}$\%$ of the binary systems are disrupted. In such cases the Be stars would be released as single stars. Moreover, the BHBe binaries that survive, characterized by extremely long orbital periods, are likely to be observationally indistinguishable from Be stars in many surveys (e.g., \citealt{Wangl2022,Carretero2023}). Consequently, our model suggests a potential formation channel for apparently single Be stars.}

\section{Discussion and Summary}
\label{sec:5}

In this letter, we present the first application of the WRLOF model to the formation of BHBe binaries, building upon the numerical simulations of \citet{ElMellah2019b}. The model's primary uncertainty stems from stellar wind parameterization, where we adopt a $\beta-$law velocity profile $\beta=2$ and simplified terminal speed prescriptions ($\beta_{\rm wind}$ and $f_{\rm w}$). While lower $\beta$ values reduce the BHBe binaries formation parameter space, they still maintain higher wind accretion efficiencies than classical BHL accretion \citep{Zuoz2021,ElMellah2019b}. We also simulated a series of binary systems with enhanced wind speeds ($f_{\rm w}=1.5,\;2,\;3$) and the results show that enhanced wind speeds ($f_{\rm w}\geq 2$) yield terminal speeds $\gtrsim 600\,\rm km\,s^{-1}$ for $M_{1}\gtrsim 50\,M_\odot$, completely suppressing Be star formation in these massive systems.  

Despite remaining uncertainties in the WRLOF model, our study reveals a potential formation pathway for Be stars in binary systems. Through population synthesis, we predict approximately {$\sim 1800-3200$ BHBe} binaries exist in our Galaxy. {The predicted numbers of BHBe binaries in the WRLOF model are approximately an order of magnitude higher than that of classical RL channel.} These systems exhibit distinctive characteristics: extremely wide orbits ($\gtrsim 1000\;\rm d$, peaking at $10000\;\rm d$) and high eccentricities, which are markedly different from those BHBe binaries formed through canonical RL channel (typically have shorter periods of $\lesssim 1000\;\rm d$). {The extremely long orbital periods of WRLOF-formed BHBe binaries, in addition to systems disrupted by BH natal kicks, suggest a potential origin for apparently single Be stars.} Our analysis excludes WRLOF formation for the recently discovered BHBe binary ALS 8814. However, Gaia's detection of wide-orbit BH binaries demonstrates the feasibility of identifying such systems through astrometric {and interferometric} methods. Future discoveries of these wide BHBe binaries will undoubtedly enhance our understanding of wind accretion processes in massive binary systems.

\begin{acknowledgements}
{We are deeply grateful to the anonymous referee for the insightful comments, which have significantly improved the quality of this work.} The authors express the gratitude to Zhaoyu Zuo and I. El Mellah for sharing the grids of wind accretion efficiencies. ZL thanks Matthias U. Kruckow for the detailed discussions about the BH formation. 
This work is supported by the Natural Science Foundation of China (grant Nos. 12288102, 12125303, 12090040/3, 12473034, 12333008, 12422305, 12273105, 12073070, 12173081), by the Strategic Priority Research Program of the Chinese Academy of Sciences (grant Nos. XDB1160303, XDB1160201, XDB1160000), the National Key R$\&$D Program of China (grant Nos. 2021YFA1600403, 2021YFA1600400), the CAS ``Light of West China", the Yunnan Revitalization Talent Support Program-Science $\&$ Technology Champion Project (No. 202305AB350003) and Young Talent project, the International Centre of Supernovae (ICESUN), Yunnan Key Laboratory of Supernova Research (Nos. 202302AN360001, 202201BC070003), Yunnan Fundamental Research Projects (No. 202401AT070139), the Natural Science Foundation of Henan Province (No. 242300420944). {X.C. acknowledges the New Cornerstone Science Foundation through the XPLORER PRIZE.}
The authors gratefully acknowledge the “PHOENIX Supercomputing Platform” jointly operated by the Binary Population Synthesis Group and the Stellar Astrophysics Group at Yunnan Observatories, Chinese Academy of Sciences. 
\end{acknowledgements}
%
%

\software{MESA (v12115; \citealt{Paxton2011,Paxton2013,Paxton2015,Paxton2018,Paxton2019})}

\newpage

\appendix

\section{{Enhanced LBV wind}}
\label{sec:A}

{Here we examine the influence of LBV winds on our results. In the wind accretion model of \citet{ElMellah2019b}, the accretion efficiency depends primarily on wind velocity rather than the mass-loss rate. Therefore, varying the mass-loss rate does not directly affect Be star formation. However, an enhanced mass-loss rate can alter stellar evolutionary tracks, thereby indirectly influencing the outcome. }

\begin{figure}[h]
    \centering
    \includegraphics[width=0.85\columnwidth]{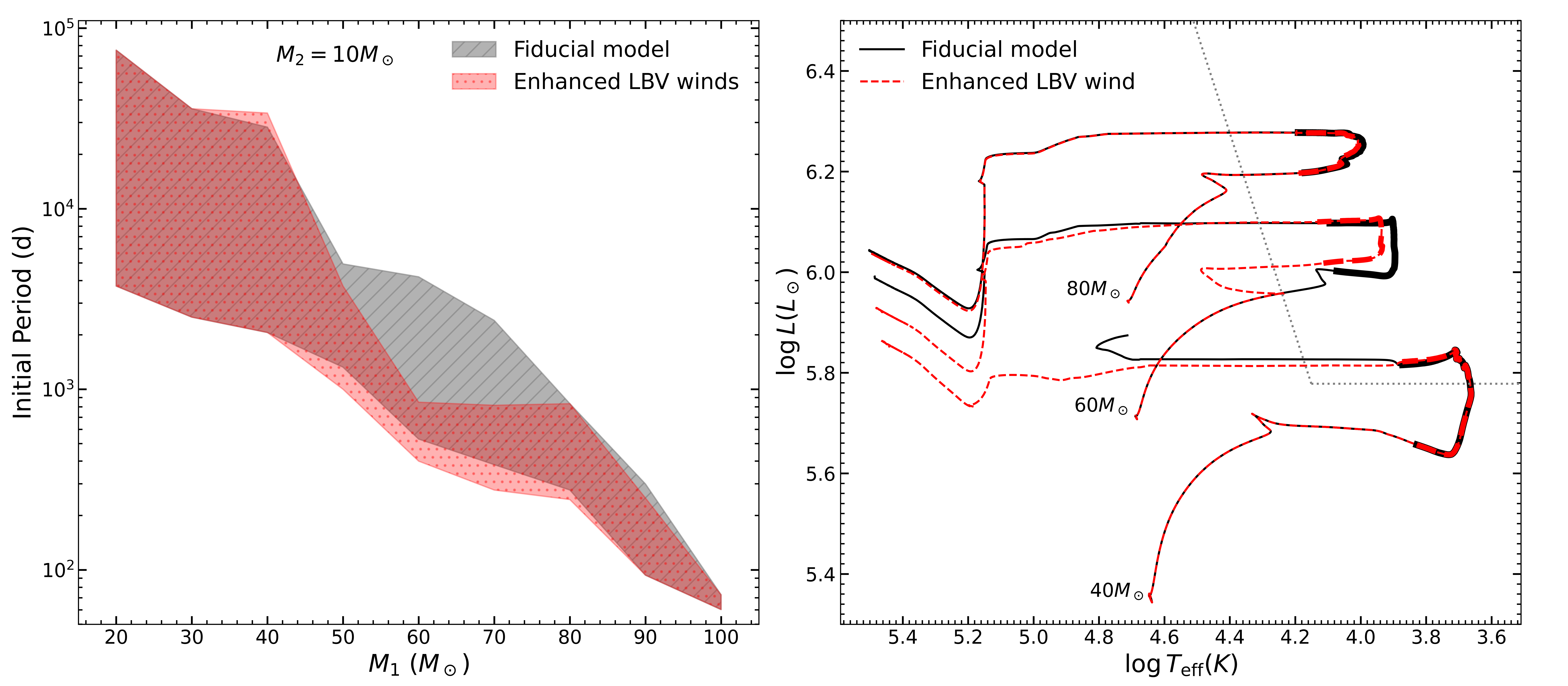}
    \caption{{Left panel: The parameter space for $M_{\rm 2}=10\,M_\odot$. The red and grey hatched regions correspond to the models with enhanced LBV winds and the fiducial case, respectively. Right panel: Comparison of several typical evolutionary tracks, where the red dashed lines and black solid lines represent the models with enhanced LBV winds and the fiducial case, respectively. The thick lines mark the wind accretion phase, defined as the period when the stellar radius lies between $0.4R_{\rm max}$ and $R_{\rm max}$, where $R_{\rm max}$ is the maximum evolutionary radius. }}
    \label{fig:A1}
\end{figure}

{To assess this effect, we recalculated a subset of grids with $M_2=10\,M_\odot$ using an enhanced LBV mass-loss rate of $10^{-4}\,M_\odot\rm yr^{-1}$ \citep{Belczynski2010}, applied when a star crosses the Humphreys-Davidson limit. The results are shown as the red hatched region in the left panel of Figure~\ref{fig:A1}, alongside the fiducial model for comparison. We find that LBV winds mainly affect systems with $M_{\rm 1}=60-70\,M_\odot$. To understand this, we plot several evolutionary tracks with different LBV wind mass-loss rates in the right panel of Figure~\ref{fig:A1}, highlighting the wind accretion phase when the star expands to between $0.4R_{\rm max}$ and $R_{\rm max}$, where $R_{\rm max}$ is the maximum evolutionary radius. For $M_1 = 60-70\,M_\odot$, the star crosses the Humphreys-Davidson limit early, prior to the formation of He core, causing the enhanced mass loss to significantly alter subsequent evolution. In contrast, lower-mass models such as the $M_1 \leq 50\,M_\odot$ and more massive systems with $M_1\geq 80\,M_\odot$ cross the Humphreys-Davidson limit after He core formation (RSG/sub-giant phase). The mass-loss rate predicted by the default wind prescription (Section~\ref{sec:2.1}) is already comparable to $10^{-4}\,M_\odot\rm yr^{-1}$. Thus, adopting an additional LBV wind prescription has limited effect in those regimes. }

{As illustrated in Figure~\ref{fig:A1}, the adoption of enhanced LBV winds reduces the parameter space for BHBe binary formation to some degree. This reduction could consequently lead to a decrease in their predicted numbers and possible adjustments to the resulting binary parameter distributions. However, as shown in Figure~\ref{fig:B1}, the relatively modest contraction in parameter space for massive primaries does not substantially alter the overall distributions of BHBe binaries.}

{It is important to note that our simplified wind model permits stellar evolution to proceed across the Humphreys-Davidson limit. Observationally, however, few stars are found beyond this limit \citep[e.g.][]{Davies2018,McDonald2022}. Various attempts have been made to explain this discrepancy \citep[e.g.][]{Cheng2024,Zapartas2025}. The issue remains under debate. While there is no doubt that adopting more complicated wind prescriptions for RSGs and LBVs would influence the parameter space of BHBe binaries, we defer such an investigation to future work.} 


\section{{Results with different wind velocities}}
\label{sec:B}

\begin{figure}[h]
    \centering
    \includegraphics[width=0.85\columnwidth]{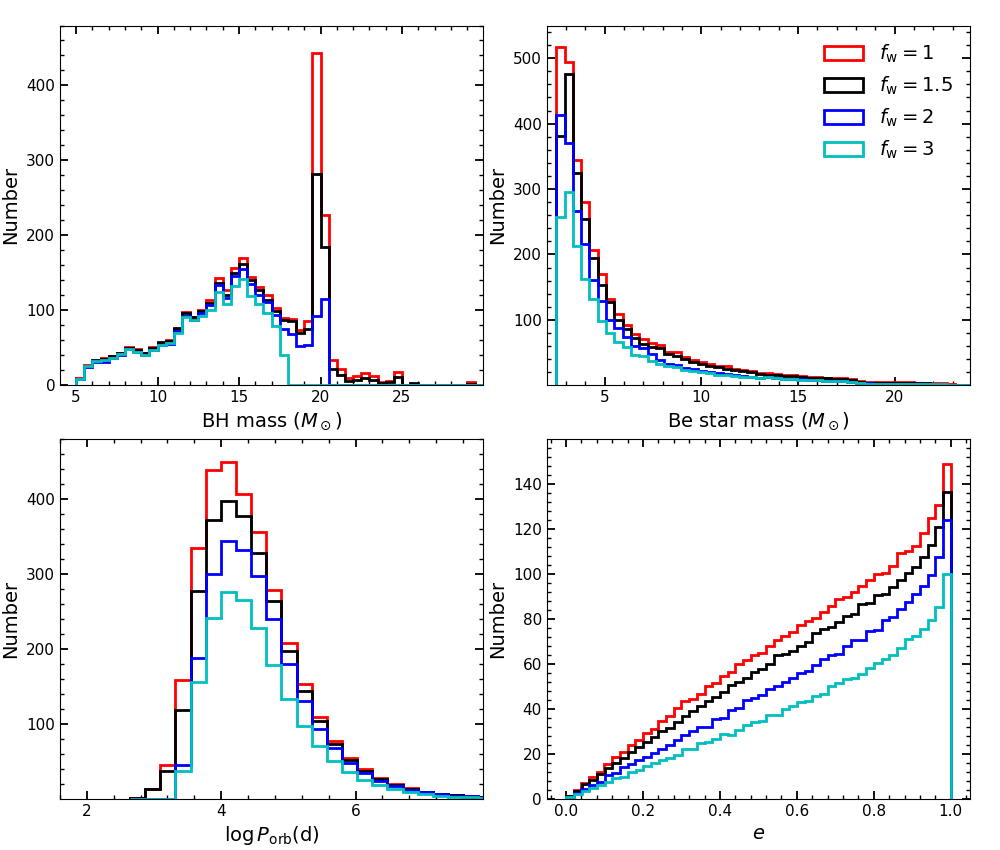}
    \caption{{Distributions of binary parameters (BH mass, Be star mass, orbital period, and eccentricity) for BHBe binaries under different wind enhancement factors $f_{\rm w}=1.5,\;2,\;3$ (corresponding to black, blue and cyan lines, respectively). The fiducial model ($f_{\rm w}=1$ is shown for comparison.)}}
    \label{fig:B1}
\end{figure}

{Figure~\ref{fig:B1} shows the statistical distributions of binary parameters in BHBe binaries for wind enhancement factors of $f_{\rm w}=1.5\,\;2,\;3$, with the fiducial case $f_{\rm w}=1$ shown for comparison. As $f_{\rm w}$ increases, the available parameter space shrinks, leading to a systematic decrease in the number of such systems. The trends of the distributions of Be star mass, orbital period, and eccentricity remain consistent; only the total number of systems decreases with increasing $f_{\rm w}$. In contrast, the BH mass distribution undergoes a noticeable change. As described in Section~\ref{sec:3.2}, higher wind velocities preferentially suppress systems with massive primaries, thereby reducing the formation of more massive BHs. } 

\section{BH mass}
\label{sec:C}

\begin{figure}[h]
    \centering
    \includegraphics[width=0.5\columnwidth]{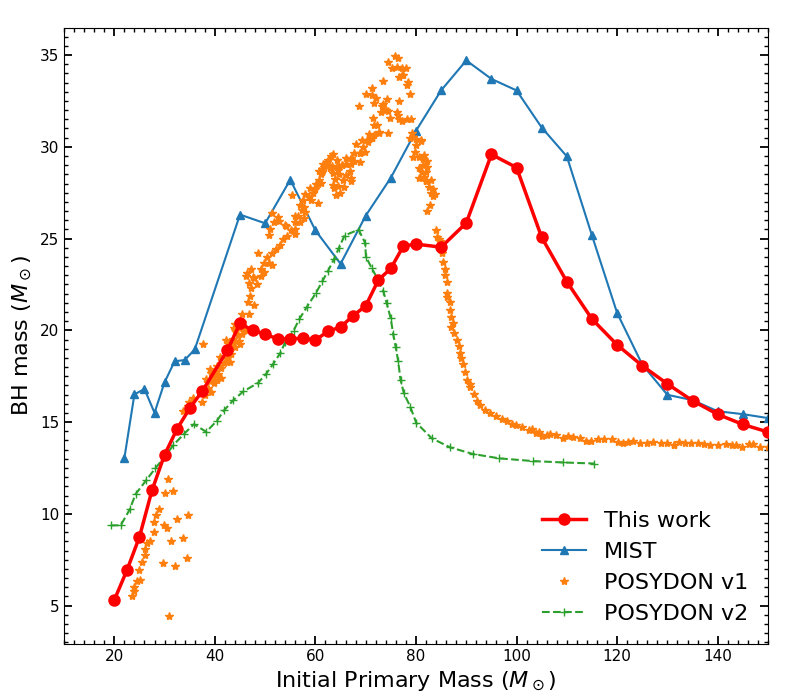}
    \caption{The final BH mass as a function initial stellar mass from different studies. The red line with circles represents this work, the blue solid line, yellow stars and greed dashed line are taken from the recent MESA simulations in other works \cite{Choi2016,Fragos2023,Bavera2023,Kruckow2024,Andrews2024}. The BH is assumed to form through direct collapse in all four models.}
    \label{fig:C1}
\end{figure}

In this work, we adopt {the delayed SN mechanism given by \citet{Fryer2012} to obtain the BH masses}. However, we emphasize that the final remnant mass for a given progenitor star exhibits significant dependence on input parameters, particularly convective overshooting and stellar wind prescriptions \citep[e.g.][]{Fragos2023,Bavera2023,Andrews2024,Kruckow2024}. Figure~\ref{fig:C1} presents comparative MESA simulations with varying input parameters, revealing substantial dispersion in the initial-final-mass relationship across different models. Due to these significant uncertainties in stellar modelling, we cannot determine which simulation is most accurate. Future observational constraints will be crucial for refining these input parameters and advancing our understanding of massive star evolution. 

\bibliography{sample701}{}

@ARTICLE{McDonald2022,
       author = {{McDonald}, Sarah L.~E. and {Davies}, Ben and {Beasor}, Emma R.},
        title = "{Red supergiants in M31: the Humphreys-Davidson limit at high metallicity}",
      journal = {\mnras},
         year = 2022,
        month = mar,
       volume = {510},
       number = {3},
        pages = {3132-3144},
          doi = {10.1093/mnras/stab3453},
archivePrefix = {arXiv},
       eprint = {2111.13716},
 primaryClass = {astro-ph.GA},
       adsurl = {https://ui.adsabs.harvard.edu/abs/2022MNRAS.510.3132M}
}

@ARTICLE{Davies2018,
       author = {{Davies}, Ben and {Crowther}, Paul A. and {Beasor}, Emma R.},
        title = "{The luminosities of cool supergiants in the Magellanic Clouds, and the Humphreys-Davidson limit revisited}",
      journal = {\mnras},
         year = 2018,
        month = aug,
       volume = {478},
       number = {3},
        pages = {3138-3148},
          doi = {10.1093/mnras/sty1302},
archivePrefix = {arXiv},
       eprint = {1804.06417},
 primaryClass = {astro-ph.SR},
       adsurl = {https://ui.adsabs.harvard.edu/abs/2018MNRAS.478.3138D}
}

@ARTICLE{Cheng2024,
       author = {{Cheng}, Shelley J. and {Goldberg}, Jared A. and {Cantiello}, Matteo and {Bauer}, Evan B. and {Renzo}, Mathieu and {Conroy}, Charlie},
        title = "{A Model for Eruptive Mass Loss in Massive Stars}",
      journal = {\apj},
         year = 2024,
        month = oct,
       volume = {974},
       number = {2},
          eid = {270},
        pages = {270},
          doi = {10.3847/1538-4357/ad701e},
archivePrefix = {arXiv},
       eprint = {2405.12274},
 primaryClass = {astro-ph.SR},
       adsurl = {https://ui.adsabs.harvard.edu/abs/2024ApJ...974..270C}
}

@ARTICLE{Zapartas2025,
       author = {{Zapartas}, E. and {de Wit}, S. and {Antoniadis}, K. and {Mu{\~n}oz-Sanchez}, G. and {Souropanis}, D. and {Bonanos}, A.~Z. and {Maravelias}, G. and {Kovlakas}, K. and {Kruckow}, M.~U. and {Fragos}, T. and {Andrews}, J.~J. and {Bavera}, S.~S. and {Briel}, M. and {Gossage}, S. and {Kasdagli}, E. and {Rocha}, K.~A. and {Sun}, M. and {Srivastava}, P.~M. and {Xing}, Z.},
        title = "{The effect of mass loss in models of red supergiants in the Small Magellanic Cloud}",
      journal = {\aap},
         year = 2025,
        month = may,
       volume = {697},
          eid = {A167},
        pages = {A167},
          doi = {10.1051/0004-6361/202452401},
archivePrefix = {arXiv},
       eprint = {2410.07335},
 primaryClass = {astro-ph.SR},
       adsurl = {https://ui.adsabs.harvard.edu/abs/2025A&A...697A.167Z}
}

@ARTICLE{Schneider2025,
       author = {{Schneider}, Fabian R.~N.},
        title = "{Theory, Simulations and Observations of Stellar Mergers}",
      journal = {arXiv e-prints},
         year = 2025,
        month = sep,
          eid = {arXiv:2509.18421},
        pages = {arXiv:2509.18421},
          doi = {10.48550/arXiv.2509.18421},
archivePrefix = {arXiv},
       eprint = {2509.18421},
 primaryClass = {astro-ph.SR},
       adsurl = {https://ui.adsabs.harvard.edu/abs/2025arXiv250918421S}
}

@ARTICLE{Belczynski2009,
       author = {{Belczynski}, Krzysztof and {Ziolkowski}, Janusz},
        title = "{On the Apparent Lack of Be X-Ray Binaries with Black Holes}",
      journal = {\apj},
         year = 2009,
        month = dec,
       volume = {707},
       number = {2},
        pages = {870-877},
          doi = {10.1088/0004-637X/707/2/870},
archivePrefix = {arXiv},
       eprint = {0907.4990},
 primaryClass = {astro-ph.GA},
       adsurl = {https://ui.adsabs.harvard.edu/abs/2009ApJ...707..870B}
}

@ARTICLE{Valli2025,
       author = {{Valli}, Ruggero and {de Mink}, Selma E. and {Justham}, Stephen and {Callister}, Thomas and {Johnston}, Cole and {Kresse}, Daniel and {Langer}, Norbert and {Rubio}, Amanda C. and {Vigna-G{\'o}mez}, Alejandro and {Wang}, Chen},
        title = "{Evidence of polar and ultralow supernova kicks from the orbits of Be X-ray binaries}",
      journal = {arXiv e-prints},
         year = 2025,
        month = may,
          eid = {arXiv:2505.08857},
        pages = {arXiv:2505.08857},
          doi = {10.48550/arXiv.2505.08857},
archivePrefix = {arXiv},
       eprint = {2505.08857},
 primaryClass = {astro-ph.HE},
       adsurl = {https://ui.adsabs.harvard.edu/abs/2025arXiv250508857V}
}

@ARTICLE{Heger2023,
       author = {{Heger}, Alexander and {M{\"u}ller}, Bernhard and {Mandel}, Ilya},
        title = "{Black holes as the end state of stellar evolution: Theory and simulations}",
      journal = {arXiv e-prints},
         year = 2023,
        month = apr,
          eid = {arXiv:2304.09350},
        pages = {arXiv:2304.09350},
          doi = {10.48550/arXiv.2304.09350},
archivePrefix = {arXiv},
       eprint = {2304.09350},
 primaryClass = {astro-ph.HE},
       adsurl = {https://ui.adsabs.harvard.edu/abs/2023arXiv230409350H}
}

@ARTICLE{Belczynski2010,
       author = {{Belczynski}, Krzysztof and {Bulik}, Tomasz and {Fryer}, Chris L. and {Ruiter}, Ashley and {Valsecchi}, Francesca and {Vink}, Jorick S. and {Hurley}, Jarrod R.},
        title = "{On the Maximum Mass of Stellar Black Holes}",
      journal = {\apj},
         year = 2010,
        month = may,
       volume = {714},
       number = {2},
        pages = {1217-1226},
          doi = {10.1088/0004-637X/714/2/1217},
archivePrefix = {arXiv},
       eprint = {0904.2784},
 primaryClass = {astro-ph.SR},
       adsurl = {https://ui.adsabs.harvard.edu/abs/2010ApJ...714.1217B}
}

@ARTICLE{Carretero2023,
       author = {{Carretero-Castrillo}, M. and {Rib{\'o}}, M. and {Paredes}, J.~M.},
        title = "{Galactic runaway O and Be stars found using Gaia DR3}",
      journal = {\aap},
         year = 2023,
        month = nov,
       volume = {679},
          eid = {A109},
        pages = {A109},
          doi = {10.1051/0004-6361/202346613},
archivePrefix = {arXiv},
       eprint = {2311.01827},
 primaryClass = {astro-ph.SR},
       adsurl = {https://ui.adsabs.harvard.edu/abs/2023A&A...679A.109C}
}

@ARTICLE{Wangl2022,
       author = {{Wang}, Luqian and {Li}, Jiao and {Wu}, You and {Gies}, Douglas R. and {Liu}, Jin Zhong and {Liu}, Chao and {Guo}, Yanjun and {Chen}, Xuefei and {Han}, Zhanwen},
        title = "{Identification of New Classical Be Stars from the LAMOST Medium Resolution Survey}",
      journal = {\apjs},
         year = 2022,
        month = jun,
       volume = {260},
       number = {2},
          eid = {35},
        pages = {35},
          doi = {10.3847/1538-4365/ac617a},
archivePrefix = {arXiv},
       eprint = {2203.15289},
 primaryClass = {astro-ph.SR},
       adsurl = {https://ui.adsabs.harvard.edu/abs/2022ApJS..260...35W}
}

@ARTICLE{Ertl2016,
       author = {{Ertl}, T. and {Janka}, H.-Th. and {Woosley}, S.~E. and {Sukhbold}, T. and {Ugliano}, M.},
        title = "{A Two-parameter Criterion for Classifying the Explodability of Massive Stars by the Neutrino-driven Mechanism}",
      journal = {\apj},
         year = 2016,
        month = feb,
       volume = {818},
       number = {2},
          eid = {124},
        pages = {124},
          doi = {10.3847/0004-637X/818/2/124},
archivePrefix = {arXiv},
       eprint = {1503.07522},
 primaryClass = {astro-ph.SR},
       adsurl = {https://ui.adsabs.harvard.edu/abs/2016ApJ...818..124E}
}

@ARTICLE{Maltsev2025,
       author = {{Maltsev}, K. and {Schneider}, F.~R.~N. and {Mandel}, I. and {M{\"u}ller}, B. and {Heger}, A. and {R{\"o}pke}, F.~K. and {Laplace}, E.},
        title = "{Explodability criteria for the neutrino-driven supernova mechanism}",
      journal = {\aap},
         year = 2025,
        month = aug,
       volume = {700},
          eid = {A20},
        pages = {A20},
          doi = {10.1051/0004-6361/202554931},
archivePrefix = {arXiv},
       eprint = {2503.23856},
 primaryClass = {astro-ph.SR},
       adsurl = {https://ui.adsabs.harvard.edu/abs/2025A&A...700A..20M}
}

@ARTICLE{Ekstrom2012,
       author = {{Ekstr{\"o}m}, S. and {Georgy}, C. and {Eggenberger}, P. and {Meynet}, G. and {Mowlavi}, N. and {Wyttenbach}, A. and {Granada}, A. and {Decressin}, T. and {Hirschi}, R. and {Frischknecht}, U. and {Charbonnel}, C. and {Maeder}, A.},
        title = "{Grids of stellar models with rotation. I. Models from 0.8 to 120 M$_{{\ensuremath{\odot}}}$ at solar metallicity (Z = 0.014)}",
      journal = {\aap},
         year = 2012,
        month = jan,
       volume = {537},
          eid = {A146},
        pages = {A146},
          doi = {10.1051/0004-6361/201117751},
archivePrefix = {arXiv},
       eprint = {1110.5049},
 primaryClass = {astro-ph.SR},
       adsurl = {https://ui.adsabs.harvard.edu/abs/2012A&A...537A.146E}
}

@ARTICLE{Elbadry2025,
       author = {{El-Badry}, Kareem and {Fabry}, Matthias and {Sana}, Hugues and {Shenar}, Tomer and {Seeburger}, Rhys},
        title = "{Complex spectral variability and hints of a luminous companion in the Be star + black hole binary candidate ALS 8814}",
      journal = {arXiv e-prints},
         year = 2025,
        month = sep,
          eid = {arXiv:2509.01545},
        pages = {arXiv:2509.01545},
          doi = {10.48550/arXiv.2509.01545},
archivePrefix = {arXiv},
       eprint = {2509.01545},
 primaryClass = {astro-ph.SR},
       adsurl = {https://ui.adsabs.harvard.edu/abs/2025arXiv250901545E}
}

@ARTICLE{Kashi2023,
       author = {{Kashi}, Amit},
        title = "{Accretion in the binary system GG Carinae and implications for B[e] supergiants}",
      journal = {\mnras},
         year = 2023,
        month = aug,
       volume = {523},
       number = {4},
        pages = {5876-5886},
          doi = {10.1093/mnras/stad1758},
archivePrefix = {arXiv},
       eprint = {2306.05992},
 primaryClass = {astro-ph.SR},
       adsurl = {https://ui.adsabs.harvard.edu/abs/2023MNRAS.523.5876K}
}

@ARTICLE{Kashi2022,
       author = {{Kashi}, Amit and {Michaelis}, Amir and {Kaminetsky}, Yarden},
        title = "{Accretion in massive colliding-wind binaries and the effect of the wind momentum ratio}",
      journal = {\mnras},
         year = 2022,
        month = nov,
       volume = {516},
       number = {3},
        pages = {3193-3205},
          doi = {10.1093/mnras/stac1912},
archivePrefix = {arXiv},
       eprint = {2207.01990},
 primaryClass = {astro-ph.SR},
       adsurl = {https://ui.adsabs.harvard.edu/abs/2022MNRAS.516.3193K}
}

@ARTICLE{Zhekov2021,
       author = {{Zhekov}, Svetozar A.},
        title = "{Colliding stellar wind modelling of the X-ray emission from WR 140}",
      journal = {\mnras},
         year = 2021,
        month = jan,
       volume = {500},
       number = {4},
        pages = {4837-4848},
          doi = {10.1093/mnras/staa3591},
archivePrefix = {arXiv},
       eprint = {2011.08550},
 primaryClass = {astro-ph.HE},
       adsurl = {https://ui.adsabs.harvard.edu/abs/2021MNRAS.500.4837Z}
}

@ARTICLE{Zhangf2004,
       author = {{Zhang}, Fan and {Li}, X. -D. and {Wang}, Z. -R.},
        title = "{Where Are the Be/Black Hole Binaries?}",
      journal = {\apj},
         year = 2004,
        month = mar,
       volume = {603},
       number = {2},
        pages = {663-668},
          doi = {10.1086/381540},
archivePrefix = {arXiv},
       eprint = {astro-ph/0311523},
 primaryClass = {astro-ph},
       adsurl = {https://ui.adsabs.harvard.edu/abs/2004ApJ...603..663Z}
}

@ARTICLE{Khokhlov2018,
       author = {{Khokhlov}, S.~A. and {Miroshnichenko}, A.~S. and {Zharikov}, S.~V. and {Manset}, N. and {Arkharov}, A.~A. and {Efimova}, N. and {Klimanov}, S. and {Larionov}, V.~M. and {Kusakin}, A.~V. and {Kokumbaeva}, R.~I. and {Omarov}, Ch. T. and {Kuratov}, K.~S. and {Kuratova}, A.~K. and {Rudy}, R.~J. and {Laag}, E.~A. and {Crawford}, K.~B. and {Swift}, T.~K. and {Puetter}, R.~C. and {Perry}, R.~B. and {Chojnowski}, S.~D. and {Agishev}, A. and {Caton}, D.~B. and {Hawkins}, R.~L. and {Smith}, A.~B. and {Reichart}, D.~E. and {Kouprianov}, V.~V. and {Haislip}, J.~B.},
        title = "{Toward Understanding the B[e] Phenomenon. VII. AS 386, a Single-lined Binary with a Candidate Black Hole Component}",
      journal = {\apj},
         year = 2018,
        month = apr,
       volume = {856},
       number = {2},
          eid = {158},
        pages = {158},
          doi = {10.3847/1538-4357/aab49d},
archivePrefix = {arXiv},
       eprint = {1803.03892},
 primaryClass = {astro-ph.SR},
       adsurl = {https://ui.adsabs.harvard.edu/abs/2018ApJ...856..158K}
}

@ARTICLE{Grudzinska2015,
       author = {{Grudzinska}, M. and {Belczynski}, K. and {Casares}, J. and {de Mink}, S.~E. and {Ziolkowski}, J. and {Negueruela}, I. and {Rib{\'o}}, M. and {Ribas}, I. and {Paredes}, J.~M. and {Herrero}, A. and {Benacquista}, M.},
        title = "{On the formation and evolution of the first Be star in a black hole binary MWC 656}",
      journal = {\mnras},
         year = 2015,
        month = sep,
       volume = {452},
       number = {3},
        pages = {2773-2787},
          doi = {10.1093/mnras/stv1419},
archivePrefix = {arXiv},
       eprint = {1504.03146},
 primaryClass = {astro-ph.HE},
       adsurl = {https://ui.adsabs.harvard.edu/abs/2015MNRAS.452.2773G}
}

@ARTICLE{Humphreys1979,
       author = {{Humphreys}, R.~M. and {Davidson}, K.},
        title = "{Studies of luminous stars in nearby galaxies. III. Comments on the evolution of the most massive stars in the Milky Way and the Large Magellanic Cloud.}",
      journal = {\apj},
         year = 1979,
        month = sep,
       volume = {232},
        pages = {409-420},
          doi = {10.1086/157301},
       adsurl = {https://ui.adsabs.harvard.edu/abs/1979ApJ...232..409H}
}

@ARTICLE{Bavera2023,
       author = {{Bavera}, Simone S. and {Fragos}, Tassos and {Zapartas}, Emmanouil and {Andrews}, Jeff J. and {Kalogera}, Vicky and {Berry}, Christopher P.~L. and {Kruckow}, Matthias and {Dotter}, Aaron and {Kovlakas}, Konstantinos and {Misra}, Devina and {Rocha}, Kyle A. and {Srivastava}, Philipp M. and {Sun}, Meng and {Xing}, Zepei},
        title = "{The formation of merging black holes with masses beyond 30 M$_{{\ensuremath{\odot}}}$ at solar metallicity}",
      journal = {Nature Astronomy},
         year = 2023,
        month = sep,
       volume = {7},
        pages = {1090-1097},
          doi = {10.1038/s41550-023-02018-5},
archivePrefix = {arXiv},
       eprint = {2212.10924},
 primaryClass = {astro-ph.HE},
       adsurl = {https://ui.adsabs.harvard.edu/abs/2023NatAs...7.1090B}
}

@ARTICLE{Wangs2024,
       author = {{Wang}, Song and {Zhao}, Xinlin and {Feng}, Fabo and {Ge}, Hongwei and {Shao}, Yong and {Cui}, Yingzhen and {Gao}, Shijie and {Zhang}, Lifu and {Wang}, Pei and {Li}, Xue and {Bai}, Zhongrui and {Yuan}, Hailong and {Huang}, Yang and {Yuan}, Haibo and {Zhang}, Zhixiang and {Yi}, Tuan and {Xiang}, Maosheng and {Li}, Zhenwei and {Li}, Tanda and {Zhang}, Junbo and {Zhang}, Meng and {Han}, Henggeng and {Fan}, Dongwei and {Li}, Xiangdong and {Chen}, Xuefei and {Liu}, Zhengwei and {Meng}, Xiangcun and {Liu}, Qingzhong and {Zhang}, Haotong and {Gu}, Wei-Min and {Liu}, Jifeng},
        title = "{A potential mass-gap black hole in a wide binary with a circular orbit}",
      journal = {Nature Astronomy},
         year = 2024,
        month = dec,
       volume = {8},
        pages = {1583-1591},
          doi = {10.1038/s41550-024-02359-9},
archivePrefix = {arXiv},
       eprint = {2409.06352},
 primaryClass = {astro-ph.SR},
       adsurl = {https://ui.adsabs.harvard.edu/abs/2024NatAs...8.1583W}
}

@ARTICLE{Andrews2022,
       author = {{Andrews}, Jeff J. and {Taggart}, Kirsty and {Foley}, Ryan},
        title = "{A Sample of Neutron Star and Black Hole Binaries Detected through Gaia DR3 Astrometry}",
      journal = {arXiv e-prints},
         year = 2022,
        month = jul,
          eid = {arXiv:2207.00680},
        pages = {arXiv:2207.00680},
          doi = {10.48550/arXiv.2207.00680},
archivePrefix = {arXiv},
       eprint = {2207.00680},
 primaryClass = {astro-ph.SR},
       adsurl = {https://ui.adsabs.harvard.edu/abs/2022arXiv220700680A}
}

@ARTICLE{Gaia2024,
       author = {{Gaia Collaboration} and {Panuzzo}, P. and {Mazeh}, T. and {Arenou}, F. and {Holl}, B. and {Caffau}, E. and {Jorissen}, A. and {Babusiaux}, C. and {Gavras}, P. and {Sahlmann}, J. and {Bastian}, U. and {Wyrzykowski}, {\L}. and {Eyer}, L. and {Leclerc}, N. and {Bauchet}, N. and {Bombrun}, A. and {Mowlavi}, N. and {Seabroke}, G.~M. and {Teyssier}, D. and {Balbinot}, E. and {Helmi}, A. and {Brown}, A.~G.~A. and {Vallenari}, A. and {Prusti}, T. and {de Bruijne}, J.~H.~J. and {Barbier}, A. and {Biermann}, M. and {Creevey}, O.~L. and {Ducourant}, C. and {Evans}, D.~W. and {Guerra}, R. and {Hutton}, A. and {Jordi}, C. and {Klioner}, S.~A. and {Lammers}, U. and {Lindegren}, L. and {Luri}, X. and {Mignard}, F. and {Nicolas}, C. and {Randich}, S. and {Sartoretti}, P. and {Smiljanic}, R. and {Tanga}, P. and {Walton}, N.~A. and {Aerts}, C. and {Bailer-Jones}, C.~A.~L. and {Cropper}, M. and {Drimmel}, R. and {Jansen}, F. and {Katz}, D. and {Lattanzi}, M.~G. and {Soubiran}, C. and {Th{\'e}venin}, F. and {van Leeuwen}, F. and {Andrae}, R. and {Audard}, M. and {Bakker}},
        title = "{Discovery of a dormant 33 solar-mass black hole in pre-release Gaia astrometry}",
      journal = {\aap},
         year = 2024,
        month = jun,
       volume = {686},
          eid = {L2},
        pages = {L2},
          doi = {10.1051/0004-6361/202449763},
archivePrefix = {arXiv},
       eprint = {2404.10486},
 primaryClass = {astro-ph.GA},
       adsurl = {https://ui.adsabs.harvard.edu/abs/2024A&A...686L...2G}
}

@ARTICLE{Tanikawa2023,
       author = {{Tanikawa}, Ataru and {Hattori}, Kohei and {Kawanaka}, Norita and {Kinugawa}, Tomoya and {Shikauchi}, Minori and {Tsuna}, Daichi},
        title = "{Search for a Black Hole Binary in Gaia DR3 Astrometric Binary Stars with Spectroscopic Data}",
      journal = {\apj},
         year = 2023,
        month = apr,
       volume = {946},
       number = {2},
          eid = {79},
        pages = {79},
          doi = {10.3847/1538-4357/acbf36},
archivePrefix = {arXiv},
       eprint = {2209.05632},
 primaryClass = {astro-ph.SR},
       adsurl = {https://ui.adsabs.harvard.edu/abs/2023ApJ...946...79T}
}

@ARTICLE{Elbadry2023b,
       author = {{El-Badry}, Kareem and {Rix}, Hans-Walter and {Cendes}, Yvette and {Rodriguez}, Antonio C. and {Conroy}, Charlie and {Quataert}, Eliot and {Hawkins}, Keith and {Zari}, Eleonora and {Hobson}, Melissa and {Breivik}, Katelyn and {Rau}, Arne and {Berger}, Edo and {Shahaf}, Sahar and {Seeburger}, Rhys and {Burdge}, Kevin B. and {Latham}, David W. and {Buchhave}, Lars A. and {Bieryla}, Allyson and {Bashi}, Dolev and {Mazeh}, Tsevi and {Faigler}, Simchon},
        title = "{A red giant orbiting a black hole}",
      journal = {\mnras},
         year = 2023,
        month = may,
       volume = {521},
       number = {3},
        pages = {4323-4348},
          doi = {10.1093/mnras/stad799},
archivePrefix = {arXiv},
       eprint = {2302.07880},
 primaryClass = {astro-ph.SR},
       adsurl = {https://ui.adsabs.harvard.edu/abs/2023MNRAS.521.4323E}
}

@ARTICLE{Elbadry2023a,
       author = {{El-Badry}, Kareem and {Rix}, Hans-Walter and {Quataert}, Eliot and {Howard}, Andrew W. and {Isaacson}, Howard and {Fuller}, Jim and {Hawkins}, Keith and {Breivik}, Katelyn and {Wong}, Kaze W.~K. and {Rodriguez}, Antonio C. and {Conroy}, Charlie and {Shahaf}, Sahar and {Mazeh}, Tsevi and {Arenou}, Fr{\'e}d{\'e}ric and {Burdge}, Kevin B. and {Bashi}, Dolev and {Faigler}, Simchon and {Weisz}, Daniel R. and {Seeburger}, Rhys and {Almada Monter}, Silvia and {Wojno}, Jennifer},
        title = "{A Sun-like star orbiting a black hole}",
      journal = {\mnras},
         year = 2023,
        month = jan,
       volume = {518},
       number = {1},
        pages = {1057-1085},
          doi = {10.1093/mnras/stac3140},
archivePrefix = {arXiv},
       eprint = {2209.06833},
 primaryClass = {astro-ph.SR},
       adsurl = {https://ui.adsabs.harvard.edu/abs/2023MNRAS.518.1057E}
}

@ARTICLE{Chakrabarti2023,
       author = {{Chakrabarti}, Sukanya and {Simon}, Joshua D. and {Craig}, Peter A. and {Reggiani}, Henrique and {Brandt}, Timothy D. and {Guhathakurta}, Puragra and {Dalba}, Paul A. and {Kirby}, Evan N. and {Chang}, Philip and {Hey}, Daniel R. and {Savino}, Alessandro and {Geha}, Marla and {Thompson}, Ian B.},
        title = "{A Noninteracting Galactic Black Hole Candidate in a Binary System with a Main-sequence Star}",
      journal = {\aj},
         year = 2023,
        month = jul,
       volume = {166},
       number = {1},
          eid = {6},
        pages = {6},
          doi = {10.3847/1538-3881/accf21},
archivePrefix = {arXiv},
       eprint = {2210.05003},
 primaryClass = {astro-ph.GA},
       adsurl = {https://ui.adsabs.harvard.edu/abs/2023AJ....166....6C}
}

@ARTICLE{Raguzova2005,
       author = {{Raguzova}, N.~V. and {Popov}, S.~B.},
        title = "{Be X-ray binaries and candidates}",
      journal = {Astronomical and Astrophysical Transactions},
         year = 2005,
        month = jun,
       volume = {24},
       number = {3},
        pages = {151-185},
          doi = {10.1080/10556790500497311},
archivePrefix = {arXiv},
       eprint = {astro-ph/0505275},
 primaryClass = {astro-ph},
       adsurl = {https://ui.adsabs.harvard.edu/abs/2005A&AT...24..151R}
}

@ARTICLE{Andrews2024,
       author = {{Andrews}, Jeff J. and {Bavera}, Simone S. and {Briel}, Max and {Chattaraj}, Abhishek and {Dotter}, Aaron and {Fragos}, Tassos and {Gallegos-Garcia}, Monica and {Gossage}, Seth and {Kalogera}, Vicky and {Kasdagli}, Eirini and {Katsaggelos}, Aggelos and {Kimball}, Chase and {Kovlakas}, Konstantinos and {Kruckow}, Matthias U. and {Liotine}, Camille and {Misra}, Devina and {Rocha}, Kyle A. and {Souropanis}, Dimitris and {Srivastava}, Philipp M. and {Sun}, Meng and {Teng}, Elizabeth and {Xing}, Zepei and {Zapartas}, Emmanouil and {Zevin}, Michael},
        title = "{POSYDON Version 2: Population Synthesis with Detailed Binary-Evolution Simulations across a Cosmological Range of Metallicities}",
      journal = {arXiv e-prints},
         year = 2024,
        month = nov,
          eid = {arXiv:2411.02376},
        pages = {arXiv:2411.02376},
          doi = {10.48550/arXiv.2411.02376},
archivePrefix = {arXiv},
       eprint = {2411.02376},
 primaryClass = {astro-ph.GA},
       adsurl = {https://ui.adsabs.harvard.edu/abs/2024arXiv241102376A}
}

@ARTICLE{Puls2008,
       author = {{Puls}, Joachim and {Vink}, Jorick S. and {Najarro}, Francisco},
        title = "{Mass loss from hot massive stars}",
      journal = {\aapr},
         year = 2008,
        month = dec,
       volume = {16},
       number = {3-4},
        pages = {209-325},
          doi = {10.1007/s00159-008-0015-8},
archivePrefix = {arXiv},
       eprint = {0811.0487},
 primaryClass = {astro-ph},
       adsurl = {https://ui.adsabs.harvard.edu/abs/2008A&ARv..16..209P}
}

@ARTICLE{Niej2017,
       author = {{Nie}, J.~D. and {Wood}, P.~R. and {Nicholls}, C.~P.},
        title = "{The Orbital Nature of 81 Ellipsoidal Red Giant Binaries in the Large Magellanic Cloud}",
      journal = {\apj},
         year = 2017,
        month = feb,
       volume = {835},
       number = {2},
          eid = {209},
        pages = {209},
          doi = {10.3847/1538-4357/835/2/209},
archivePrefix = {arXiv},
       eprint = {1702.02376},
 primaryClass = {astro-ph.SR},
       adsurl = {https://ui.adsabs.harvard.edu/abs/2017ApJ...835..209N}
}

@ARTICLE{Fragos2023,
       author = {{Fragos}, Tassos and {Andrews}, Jeff J. and {Bavera}, Simone S. and {Berry}, Christopher P.~L. and {Coughlin}, Scott and {Dotter}, Aaron and {Giri}, Prabin and {Kalogera}, Vicky and {Katsaggelos}, Aggelos and {Kovlakas}, Konstantinos and {Lalvani}, Shamal and {Misra}, Devina and {Srivastava}, Philipp M. and {Qin}, Ying and {Rocha}, Kyle A. and {Rom{\'a}n-Garza}, Jaime and {Serra}, Juan Gabriel and {Stahle}, Petter and {Sun}, Meng and {Teng}, Xu and {Trajcevski}, Goce and {Tran}, Nam Hai and {Xing}, Zepei and {Zapartas}, Emmanouil and {Zevin}, Michael},
        title = "{POSYDON: A General-purpose Population Synthesis Code with Detailed Binary-evolution Simulations}",
      journal = {\apjs},
         year = 2023,
        month = feb,
       volume = {264},
       number = {2},
          eid = {45},
        pages = {45},
          doi = {10.3847/1538-4365/ac90c1},
archivePrefix = {arXiv},
       eprint = {2202.05892},
 primaryClass = {astro-ph.SR},
       adsurl = {https://ui.adsabs.harvard.edu/abs/2023ApJS..264...45F}
}

@ARTICLE{Kotko2024,
       author = {{Kotko}, I. and {Banerjee}, S. and {Belczynski}, K.},
        title = "{The enigmatic origin of two dormant BH binaries: Gaia BH1 and Gaia BH2}",
      journal = {\mnras},
         year = 2024,
        month = dec,
       volume = {535},
       number = {4},
        pages = {3577-3594},
          doi = {10.1093/mnras/stae2591},
archivePrefix = {arXiv},
       eprint = {2403.13579},
 primaryClass = {astro-ph.SR},
       adsurl = {https://ui.adsabs.harvard.edu/abs/2024MNRAS.535.3577K}
}

@ARTICLE{Ducci2009,
       author = {{Ducci}, L. and {Sidoli}, L. and {Mereghetti}, S. and {Paizis}, A. and {Romano}, P.},
        title = "{The structure of blue supergiant winds and the accretion in supergiant high-mass X-ray binaries}",
      journal = {\mnras},
         year = 2009,
        month = oct,
       volume = {398},
       number = {4},
        pages = {2152-2165},
          doi = {10.1111/j.1365-2966.2009.15265.x},
archivePrefix = {arXiv},
       eprint = {0906.3185},
 primaryClass = {astro-ph.HE},
       adsurl = {https://ui.adsabs.harvard.edu/abs/2009MNRAS.398.2152D}
}

@ARTICLE{ElMellah2018,
       author = {{El Mellah}, I. and {Sundqvist}, J.~O. and {Keppens}, R.},
        title = "{Accretion from a clumpy massive-star wind in supergiant X-ray binaries}",
      journal = {\mnras},
         year = 2018,
        month = apr,
       volume = {475},
       number = {3},
        pages = {3240-3252},
          doi = {10.1093/mnras/stx3211},
archivePrefix = {arXiv},
       eprint = {1711.08709},
 primaryClass = {astro-ph.HE},
       adsurl = {https://ui.adsabs.harvard.edu/abs/2018MNRAS.475.3240E}
}

@ARTICLE{ElMellah2017,
       author = {{El Mellah}, I. and {Casse}, F.},
        title = "{A numerical investigation of wind accretion in persistent supergiant X-ray binaries - I. Structure of the flow at the orbital scale}",
      journal = {\mnras},
         year = 2017,
        month = may,
       volume = {467},
       number = {3},
        pages = {2585-2593},
          doi = {10.1093/mnras/stx225},
archivePrefix = {arXiv},
       eprint = {1609.01532},
 primaryClass = {astro-ph.HE},
       adsurl = {https://ui.adsabs.harvard.edu/abs/2017MNRAS.467.2585E}
}

@ARTICLE{Choi2016,
       author = {{Choi}, Jieun and {Dotter}, Aaron and {Conroy}, Charlie and {Cantiello}, Matteo and {Paxton}, Bill and {Johnson}, Benjamin D.},
        title = "{Mesa Isochrones and Stellar Tracks (MIST). I. Solar-scaled Models}",
      journal = {\apj},
         year = 2016,
        month = jun,
       volume = {823},
       number = {2},
          eid = {102},
        pages = {102},
          doi = {10.3847/0004-637X/823/2/102},
archivePrefix = {arXiv},
       eprint = {1604.08592},
 primaryClass = {astro-ph.SR},
       adsurl = {https://ui.adsabs.harvard.edu/abs/2016ApJ...823..102C}
}

@ARTICLE{Vink2001,
       author = {{Vink}, Jorick S. and {de Koter}, A. and {Lamers}, H.~J.~G.~L.~M.},
        title = "{Mass-loss predictions for O and B stars as a function of metallicity}",
      journal = {\aap},
         year = 2001,
        month = apr,
       volume = {369},
        pages = {574-588},
          doi = {10.1051/0004-6361:20010127},
archivePrefix = {arXiv},
       eprint = {astro-ph/0101509},
 primaryClass = {astro-ph},
       adsurl = {https://ui.adsabs.harvard.edu/abs/2001A&A...369..574V}
}

@ARTICLE{Kruckow2024,
       author = {{Kruckow}, Matthias U. and {Andrews}, Jeff J. and {Fragos}, Tassos and {Holl}, Berry and {Bavera}, Simone S. and {Briel}, Max and {Gossage}, Seth and {Kovlakas}, Konstantinos and {Rocha}, Kyle A. and {Sun}, Meng and {Srivastava}, Philipp M. and {Xing}, Zepei and {Zapartas}, Emmanouil},
        title = "{The formation of black holes in non-interacting isolated binaries: Gaia black holes as calibrators of stellar winds from massive stars}",
      journal = {\aap},
         year = 2024,
        month = dec,
       volume = {692},
          eid = {A141},
        pages = {A141},
          doi = {10.1051/0004-6361/202452356},
archivePrefix = {arXiv},
       eprint = {2410.18501},
 primaryClass = {astro-ph.SR},
       adsurl = {https://ui.adsabs.harvard.edu/abs/2024A&A...692A.141K}
}

@ARTICLE{Hobbs2005,
       author = {{Hobbs}, G. and {Lorimer}, D.~R. and {Lyne}, A.~G. and {Kramer}, M.},
        title = "{A statistical study of 233 pulsar proper motions}",
      journal = {\mnras},
         year = 2005,
        month = jul,
       volume = {360},
       number = {3},
        pages = {974-992},
          doi = {10.1111/j.1365-2966.2005.09087.x},
archivePrefix = {arXiv},
       eprint = {astro-ph/0504584},
 primaryClass = {astro-ph},
       adsurl = {https://ui.adsabs.harvard.edu/abs/2005MNRAS.360..974H}
}

@ARTICLE{Fryer2012,
       author = {{Fryer}, Chris L. and {Belczynski}, Krzysztof and {Wiktorowicz}, Grzegorz and {Dominik}, Michal and {Kalogera}, Vicky and {Holz}, Daniel E.},
        title = "{Compact Remnant Mass Function: Dependence on the Explosion Mechanism and Metallicity}",
      journal = {\apj},
         year = 2012,
        month = apr,
       volume = {749},
       number = {1},
          eid = {91},
        pages = {91},
          doi = {10.1088/0004-637X/749/1/91},
archivePrefix = {arXiv},
       eprint = {1110.1726},
 primaryClass = {astro-ph.SR},
       adsurl = {https://ui.adsabs.harvard.edu/abs/2012ApJ...749...91F}
}

@ARTICLE{Sen2024,
       author = {{Sen}, K. and {El Mellah}, I. and {Langer}, N. and {Xu}, X. -T. and {Quast}, M. and {Pauli}, D.},
        title = "{Whispering in the dark: Faint X-ray emission from black holes with OB star companions}",
      journal = {\aap},
         year = 2024,
        month = oct,
       volume = {690},
          eid = {A256},
        pages = {A256},
          doi = {10.1051/0004-6361/202450940},
archivePrefix = {arXiv},
       eprint = {2406.08596},
 primaryClass = {astro-ph.HE},
       adsurl = {https://ui.adsabs.harvard.edu/abs/2024A&A...690A.256S}
}

@ARTICLE{Petrovic2005,
       author = {{Petrovic}, J. and {Langer}, N. and {van der Hucht}, K.~A.},
        title = "{Constraining the mass transfer in massive binaries through progenitor evolution models of Wolf-Rayet+O binaries}",
      journal = {\aap},
         year = 2005,
        month = jun,
       volume = {435},
       number = {3},
        pages = {1013-1030},
          doi = {10.1051/0004-6361:20042368},
archivePrefix = {arXiv},
       eprint = {astro-ph/0504242},
 primaryClass = {astro-ph},
       adsurl = {https://ui.adsabs.harvard.edu/abs/2005A&A...435.1013P}
}

@ARTICLE{deMink2013,
       author = {{de Mink}, S.~E. and {Langer}, N. and {Izzard}, R.~G. and {Sana}, H. and {de Koter}, A.},
        title = "{The Rotation Rates of Massive Stars: The Role of Binary Interaction through Tides, Mass Transfer, and Mergers}",
      journal = {\apj},
         year = 2013,
        month = feb,
       volume = {764},
       number = {2},
          eid = {166},
        pages = {166},
          doi = {10.1088/0004-637X/764/2/166},
archivePrefix = {arXiv},
       eprint = {1211.3742},
 primaryClass = {astro-ph.SR},
       adsurl = {https://ui.adsabs.harvard.edu/abs/2013ApJ...764..166D}
}

@ARTICLE{Belczynski2008,
       author = {{Belczynski}, Krzysztof and {Kalogera}, Vassiliki and {Rasio}, Frederic A. and {Taam}, Ronald E. and {Zezas}, Andreas and {Bulik}, Tomasz and {Maccarone}, Thomas J. and {Ivanova}, Natalia},
        title = "{Compact Object Modeling with the StarTrack Population Synthesis Code}",
      journal = {\apjs},
         year = 2008,
        month = jan,
       volume = {174},
       number = {1},
        pages = {223-260},
          doi = {10.1086/521026},
archivePrefix = {arXiv},
       eprint = {astro-ph/0511811},
 primaryClass = {astro-ph},
       adsurl = {https://ui.adsabs.harvard.edu/abs/2008ApJS..174..223B}
}

@ARTICLE{Vink2022,
       author = {{Vink}, Jorick S.},
        title = "{Theory and Diagnostics of Hot Star Mass Loss}",
      journal = {\araa},
         year = 2022,
        month = aug,
       volume = {60},
        pages = {203-246},
          doi = {10.1146/annurev-astro-052920-094949},
archivePrefix = {arXiv},
       eprint = {2109.08164},
 primaryClass = {astro-ph.SR},
       adsurl = {https://ui.adsabs.harvard.edu/abs/2022ARA&A..60..203V}
}

@ARTICLE{Sander2017,
       author = {{Sander}, A.~A.~C. and {Hamann}, W. -R. and {Todt}, H. and {Hainich}, R. and {Shenar}, T.},
        title = "{Coupling hydrodynamics with comoving frame radiative transfer. I. A unified approach for OB and WR stars}",
      journal = {\aap},
         year = 2017,
        month = jul,
       volume = {603},
          eid = {A86},
        pages = {A86},
          doi = {10.1051/0004-6361/201730642},
archivePrefix = {arXiv},
       eprint = {1704.08698},
 primaryClass = {astro-ph.SR},
       adsurl = {https://ui.adsabs.harvard.edu/abs/2017A&A...603A..86S}
}

@ARTICLE{Gimenez2016,
       author = {{Gim{\'e}nez-Garc{\'\i}a}, A. and {Shenar}, T. and {Torrej{\'o}n}, J.~M. and {Oskinova}, L. and {Mart{\'\i}nez-N{\'u}{\~n}ez}, S. and {Hamann}, W. -R. and {Rodes-Roca}, J.~J. and {Gonz{\'a}lez-Gal{\'a}n}, A. and {Alonso-Santiago}, J. and {Gonz{\'a}lez-Fern{\'a}ndez}, C. and {Bernabeu}, G. and {Sander}, A.},
        title = "{Measuring the stellar wind parameters in IGR J17544-2619 and Vela X-1 constrains the accretion physics in supergiant fast X-ray transient and classical supergiant X-ray binaries}",
      journal = {\aap},
         year = 2016,
        month = jun,
       volume = {591},
          eid = {A26},
        pages = {A26},
          doi = {10.1051/0004-6361/201527551},
archivePrefix = {arXiv},
       eprint = {1603.00925},
 primaryClass = {astro-ph.HE},
       adsurl = {https://ui.adsabs.harvard.edu/abs/2016A&A...591A..26G}
}

@ARTICLE{Hamann1995,
       author = {{Hamann}, W. -R. and {Koesterke}, L. and {Wessolowski}, U.},
        title = "{Spectral analyses of the Galactic Wolf-Rayet stars: hydrogen-helium abundances and improved stellar parameters for the WN class}",
      journal = {\aap},
         year = 1995,
        month = jul,
       volume = {299},
        pages = {151},
       adsurl = {https://ui.adsabs.harvard.edu/abs/1995A&A...299..151H}
}

@ARTICLE{Yoon2006,
       author = {{Yoon}, S. -C. and {Langer}, N. and {Norman}, C.},
        title = "{Single star progenitors of long gamma-ray bursts. I. Model grids and redshift dependent GRB rate}",
      journal = {\aap},
         year = 2006,
        month = dec,
       volume = {460},
       number = {1},
        pages = {199-208},
          doi = {10.1051/0004-6361:20065912},
archivePrefix = {arXiv},
       eprint = {astro-ph/0606637},
 primaryClass = {astro-ph},
       adsurl = {https://ui.adsabs.harvard.edu/abs/2006A&A...460..199Y}
}

@ARTICLE{Brott2011,
       author = {{Brott}, I. and {de Mink}, S.~E. and {Cantiello}, M. and {Langer}, N. and {de Koter}, A. and {Evans}, C.~J. and {Hunter}, I. and {Trundle}, C. and {Vink}, J.~S.},
        title = "{Rotating massive main-sequence stars. I. Grids of evolutionary models and isochrones}",
      journal = {\aap},
         year = 2011,
        month = jun,
       volume = {530},
          eid = {A115},
        pages = {A115},
          doi = {10.1051/0004-6361/201016113},
archivePrefix = {arXiv},
       eprint = {1102.0530},
 primaryClass = {astro-ph.SR},
       adsurl = {https://ui.adsabs.harvard.edu/abs/2011A&A...530A.115B}
}

@ARTICLE{Henyey1965,
       author = {{Henyey}, Louis and {Vardya}, M.~S. and {Bodenheimer}, Peter},
        title = "{Studies in Stellar Evolution. III. The Calculation of Model Envelopes.}",
      journal = {\apj},
         year = 1965,
        month = oct,
       volume = {142},
        pages = {841},
          doi = {10.1086/148357},
       adsurl = {https://ui.adsabs.harvard.edu/abs/1965ApJ...142..841H}
}

@ARTICLE{Iglesias1996,
       author = {{Iglesias}, Carlos A. and {Rogers}, Forrest J.},
        title = "{Updated Opal Opacities}",
      journal = {\apj},
         year = 1996,
        month = jun,
       volume = {464},
        pages = {943},
          doi = {10.1086/177381},
       adsurl = {https://ui.adsabs.harvard.edu/abs/1996ApJ...464..943I}
}

@ARTICLE{Asplund2009,
       author = {{Asplund}, Martin and {Grevesse}, Nicolas and {Sauval}, A. Jacques and {Scott}, Pat},
        title = "{The Chemical Composition of the Sun}",
      journal = {\araa},
         year = 2009,
        month = sep,
       volume = {47},
       number = {1},
        pages = {481-522},
          doi = {10.1146/annurev.astro.46.060407.145222},
archivePrefix = {arXiv},
       eprint = {0909.0948},
 primaryClass = {astro-ph.SR},
       adsurl = {https://ui.adsabs.harvard.edu/abs/2009ARA&A..47..481A}
}

@ARTICLE{Janssens2023,
       author = {{Janssens}, S. and {Shenar}, T. and {Degenaar}, N. and {Bodensteiner}, J. and {Sana}, H. and {Audenaert}, J. and {Frost}, A.~J.},
        title = "{MWC 656 is unlikely to contain a black hole}",
      journal = {\aap},
         year = 2023,
        month = sep,
       volume = {677},
          eid = {L9},
        pages = {L9},
          doi = {10.1051/0004-6361/202347318},
archivePrefix = {arXiv},
       eprint = {2308.08642},
 primaryClass = {astro-ph.SR},
       adsurl = {https://ui.adsabs.harvard.edu/abs/2023A&A...677L...9J}
}

@INPROCEEDINGS{Rivinius2024b,
       author = {{Rivinius}, Th. and {Klement}, R. and {Chojnowski}, S.~D. and {Baade}, D. and {Shepard}, K. and {Hadrava}, P.},
        title = "{MWC 656: A Be+BH or a Be+sdO?}",
    booktitle = {Massive Stars Near and Far},
         year = 2024,
       editor = {{Mackey}, Jonathan and {Vink}, Jorick S. and {St-Louis}, Nicode},
       series = {IAU Symposium},
       volume = {361},
        month = jan,
        pages = {332-333},
          doi = {10.1017/S1743921322002976},
archivePrefix = {arXiv},
       eprint = {2208.12315},
 primaryClass = {astro-ph.SR},
       adsurl = {https://ui.adsabs.harvard.edu/abs/2024IAUS..361..332R}
}

@ARTICLE{An2025,
       author = {{An}, Qian-Yu and {Huang}, Yang and {Gu}, Wei-Min and {Shao}, Yong and {Zhang}, Zhi-Xiang and {Yi}, Tuan and {Lailey}, B.~D. and {Sigut}, T.~A.~A. and {Akira Rocha}, Kyle and {Sun}, Meng and {Gossage}, Seth and {Gao}, Shi-Jie and {Weng}, Shan-Shan and {Wang}, Song and {Zhang}, Bowen and {Zhao}, Xinlin and {Qi}, Senyu and {Liao}, Shilong and {Ji}, Jianghui and {Wang}, Junfeng and {Wu}, Jianfeng and {Sun}, Mouyuan and {Li}, Xiang-Dong and {Liu}, Jifeng},
        title = "{A Be star-black hole binary with a wide orbit from LAMOST time-domain survey}",
      journal = {arXiv e-prints},
         year = 2025,
        month = may,
          eid = {arXiv:2505.23151},
        pages = {arXiv:2505.23151},
          doi = {10.48550/arXiv.2505.23151},
archivePrefix = {arXiv},
       eprint = {2505.23151},
 primaryClass = {astro-ph.SR},
       adsurl = {https://ui.adsabs.harvard.edu/abs/2025arXiv250523151A}
}

@ARTICLE{Casares2014,
       author = {{Casares}, J. and {Negueruela}, I. and {Rib{\'o}}, M. and {Ribas}, I. and {Paredes}, J.~M. and {Herrero}, A. and {Sim{\'o}n-D{\'\i}az}, S.},
        title = "{A Be-type star with a black-hole companion}",
      journal = {\nat},
         year = 2014,
        month = jan,
       volume = {505},
       number = {7483},
        pages = {378-381},
          doi = {10.1038/nature12916},
archivePrefix = {arXiv},
       eprint = {1401.3711},
 primaryClass = {astro-ph.SR},
       adsurl = {https://ui.adsabs.harvard.edu/abs/2014Natur.505..378C}
}

@ARTICLE{Wiktorowicz2021,
       author = {{Wiktorowicz}, Grzegorz and {Lasota}, Jean-Pierre and {Belczynski}, Krzysztof and {Lu}, Youjun and {Liu}, Jifeng and {I{\l}kiewicz}, Krystian},
        title = "{Wind-powered Ultraluminous X-ray Sources}",
      journal = {\apj},
         year = 2021,
        month = sep,
       volume = {918},
       number = {2},
          eid = {60},
        pages = {60},
          doi = {10.3847/1538-4357/ac0cf7},
archivePrefix = {arXiv},
       eprint = {2103.02026},
 primaryClass = {astro-ph.HE},
       adsurl = {https://ui.adsabs.harvard.edu/abs/2021ApJ...918...60W}
}

@ARTICLE{Zuoz2021,
       author = {{Zuo}, Zhao-Yu and {Song}, Hao-Tian and {Xue}, Han-Chen},
        title = "{Population synthesis on ultra-luminous X-ray sources with an accreting neutron star: Wind Roche-lobe overflow cases}",
      journal = {\aap},
         year = 2021,
        month = may,
       volume = {649},
          eid = {L2},
        pages = {L2},
          doi = {10.1051/0004-6361/202140792},
       adsurl = {https://ui.adsabs.harvard.edu/abs/2021A&A...649L...2Z}
}

@ARTICLE{ElMellah2019a,
       author = {{El Mellah}, I. and {Sander}, A.~A.~C. and {Sundqvist}, J.~O. and {Keppens}, R.},
        title = "{Formation of wind-captured disks in supergiant X-ray binaries. Consequences for Vela X-1 and Cygnus X-1}",
      journal = {\aap},
         year = 2019,
        month = feb,
       volume = {622},
          eid = {A189},
        pages = {A189},
          doi = {10.1051/0004-6361/201834498},
archivePrefix = {arXiv},
       eprint = {1810.12933},
 primaryClass = {astro-ph.HE},
       adsurl = {https://ui.adsabs.harvard.edu/abs/2019A&A...622A.189E}
}

@ARTICLE{ElMellah2019b,
       author = {{El Mellah}, I. and {Sundqvist}, J.~O. and {Keppens}, R.},
        title = "{Wind Roche lobe overflow in high-mass X-ray binaries. A possible mass-transfer mechanism for ultraluminous X-ray sources}",
      journal = {\aap},
         year = 2019,
        month = feb,
       volume = {622},
          eid = {L3},
        pages = {L3},
          doi = {10.1051/0004-6361/201834543},
archivePrefix = {arXiv},
       eprint = {1810.12937},
 primaryClass = {astro-ph.HE},
       adsurl = {https://ui.adsabs.harvard.edu/abs/2019A&A...622L...3E}
}

@ARTICLE{Skopal2023,
       author = {{Skopal}, A. and {Shagatova}, N.},
        title = "{Wind-mass transfer in S-type symbiotic binaries. IV. Indication of high wind-mass-transfer efficiency from active phases}",
      journal = {\aap},
         year = 2023,
        month = dec,
       volume = {680},
          eid = {A60},
        pages = {A60},
          doi = {10.1051/0004-6361/202347396},
archivePrefix = {arXiv},
       eprint = {2310.13133},
 primaryClass = {astro-ph.SR},
       adsurl = {https://ui.adsabs.harvard.edu/abs/2023A&A...680A..60S}
}

@ARTICLE{Shagatova2021,
       author = {{Shagatova}, N. and {Skopal}, A. and {Shugarov}, S. Yu. and {Kom{\v{z}}{\'\i}k}, R. and {Kundra}, E. and {Teyssier}, F.},
        title = "{Wind mass transfer in S-type symbiotic binaries. III. Confirmation of a wind focusing in EG Andromedae from the nebular [O III] {\ensuremath{\lambda}}5007 line}",
      journal = {\aap},
         year = 2021,
        month = feb,
       volume = {646},
          eid = {A116},
        pages = {A116},
          doi = {10.1051/0004-6361/202039103},
archivePrefix = {arXiv},
       eprint = {2012.08417},
 primaryClass = {astro-ph.SR},
       adsurl = {https://ui.adsabs.harvard.edu/abs/2021A&A...646A.116S}
}

@ARTICLE{Skopal2015,
       author = {{Skopal}, A. and {Carikov{\'a}}, Z.},
        title = "{Wind mass transfer in S-type symbiotic binaries. I. Focusing by the wind compression model}",
      journal = {\aap},
         year = 2015,
        month = jan,
       volume = {573},
          eid = {A8},
        pages = {A8},
          doi = {10.1051/0004-6361/201424779},
archivePrefix = {arXiv},
       eprint = {1410.7674},
 primaryClass = {astro-ph.SR},
       adsurl = {https://ui.adsabs.harvard.edu/abs/2015A&A...573A...8S}
}

@ARTICLE{Shagatova2016,
       author = {{Shagatova}, N. and {Skopal}, A. and {Carikov{\'a}}, Z.},
        title = "{Wind mass transfer in S-type symbiotic binaries. II. Indication of wind focusing}",
      journal = {\aap},
         year = 2016,
        month = apr,
       volume = {588},
          eid = {A83},
        pages = {A83},
          doi = {10.1051/0004-6361/201525645},
archivePrefix = {arXiv},
       eprint = {1602.04640},
 primaryClass = {astro-ph.SR},
       adsurl = {https://ui.adsabs.harvard.edu/abs/2016A&A...588A..83S}
}

@ARTICLE{Saladino2018,
       author = {{Saladino}, M.~I. and {Pols}, O.~R. and {van der Helm}, E. and {Pelupessy}, I. and {Portegies Zwart}, S.},
        title = "{Gone with the wind: the impact of wind mass transfer on the orbital evolution of AGB binary systems}",
      journal = {\aap},
         year = 2018,
        month = oct,
       volume = {618},
          eid = {A50},
        pages = {A50},
          doi = {10.1051/0004-6361/201832967},
archivePrefix = {arXiv},
       eprint = {1805.03208},
 primaryClass = {astro-ph.SR},
       adsurl = {https://ui.adsabs.harvard.edu/abs/2018A&A...618A..50S}
}

@ARTICLE{Saladino2019,
       author = {{Saladino}, M.~I. and {Pols}, O.~R. and {Abate}, C.},
        title = "{Slowly, slowly in the wind. 3D hydrodynamical simulations of wind mass transfer and angular-momentum loss in AGB binary systems}",
      journal = {\aap},
         year = 2019,
        month = jun,
       volume = {626},
          eid = {A68},
        pages = {A68},
          doi = {10.1051/0004-6361/201834598},
archivePrefix = {arXiv},
       eprint = {1903.04515},
 primaryClass = {astro-ph.SR},
       adsurl = {https://ui.adsabs.harvard.edu/abs/2019A&A...626A..68S}
}

@ARTICLE{Chenz2017,
       author = {{Chen}, Zhuo and {Frank}, Adam and {Blackman}, Eric G. and {Nordhaus}, Jason and {Carroll-Nellenback}, Jonathan},
        title = "{Mass transfer and disc formation in AGB binary systems}",
      journal = {\mnras},
         year = 2017,
        month = jul,
       volume = {468},
       number = {4},
        pages = {4465-4477},
          doi = {10.1093/mnras/stx680},
archivePrefix = {arXiv},
       eprint = {1702.06160},
 primaryClass = {astro-ph.SR},
       adsurl = {https://ui.adsabs.harvard.edu/abs/2017MNRAS.468.4465C}
}

@ARTICLE{Liuz2017,
       author = {{Liu}, Zheng-Wei and {Stancliffe}, Richard J. and {Abate}, Carlo and {Matrozis}, Elvijs},
        title = "{Three-dimensional Hydrodynamical Simulations of Mass Transfer in Binary Systems by a Free Wind}",
      journal = {\apj},
         year = 2017,
        month = sep,
       volume = {846},
       number = {2},
          eid = {117},
        pages = {117},
          doi = {10.3847/1538-4357/aa8622},
archivePrefix = {arXiv},
       eprint = {1708.03639},
 primaryClass = {astro-ph.SR},
       adsurl = {https://ui.adsabs.harvard.edu/abs/2017ApJ...846..117L}
}

@ARTICLE{Mohamed2012,
       author = {{Mohamed}, S. and {Podsiadlowski}, Ph.},
        title = "{Mass Transfer in Mira-type Binaries}",
      journal = {Baltic Astronomy},
         year = 2012,
        month = jan,
       volume = {21},
        pages = {88-96},
          doi = {10.1515/astro-2017-0362},
       adsurl = {https://ui.adsabs.harvard.edu/abs/2012BaltA..21...88M}
}

@ARTICLE{Abate2013,
       author = {{Abate}, C. and {Pols}, O.~R. and {Izzard}, R.~G. and {Mohamed}, S.~S. and {de Mink}, S.~E.},
        title = "{Wind Roche-lobe overflow: Application to carbon-enhanced metal-poor stars}",
      journal = {\aap},
         year = 2013,
        month = apr,
       volume = {552},
          eid = {A26},
        pages = {A26},
          doi = {10.1051/0004-6361/201220007},
archivePrefix = {arXiv},
       eprint = {1302.4441},
 primaryClass = {astro-ph.SR},
       adsurl = {https://ui.adsabs.harvard.edu/abs/2013A&A...552A..26A}
}

@ARTICLE{Bondi1944,
       author = {{Bondi}, H. and {Hoyle}, F.},
        title = "{On the mechanism of accretion by stars}",
      journal = {\mnras},
         year = 1944,
        month = jan,
       volume = {104},
        pages = {273},
          doi = {10.1093/mnras/104.5.273},
       adsurl = {https://ui.adsabs.harvard.edu/abs/1944MNRAS.104..273B}
}

@ARTICLE{Hoyle1939,
       author = {{Hoyle}, F. and {Lyttleton}, R.~A.},
        title = "{The effect of interstellar matter on climatic variation}",
      journal = {Proceedings of the Cambridge Philosophical Society},
         year = 1939,
        month = jan,
       volume = {35},
       number = {3},
        pages = {405},
          doi = {10.1017/S0305004100021150},
       adsurl = {https://ui.adsabs.harvard.edu/abs/1939PCPS...35..405H}
}

@ARTICLE{Chenx2024,
       author = {{Chen}, Xuefei and {Liu}, Zhengwei and {Han}, Zhanwen},
        title = "{Binary stars in the new millennium}",
      journal = {Progress in Particle and Nuclear Physics},
         year = 2024,
        month = jan,
       volume = {134},
          eid = {104083},
        pages = {104083},
          doi = {10.1016/j.ppnp.2023.104083},
archivePrefix = {arXiv},
       eprint = {2311.11454},
 primaryClass = {astro-ph.SR},
       adsurl = {https://ui.adsabs.harvard.edu/abs/2024PrPNP.13404083C}
}

@ARTICLE{Sunm2024,
       author = {{Sun}, Meng and {Levina}, Sasha and {Gossage}, Seth and {Kalogera}, Vicky and {Leiner}, Emily M. and {Geller}, Aaron M. and {Doctor}, Zoheyr},
        title = "{Wind Roche-lobe Overflow in Low-mass Binaries: Exploring the Origin of Rapidly Rotating Blue Lurkers}",
      journal = {\apj},
         year = 2024,
        month = jul,
       volume = {969},
       number = {1},
          eid = {8},
        pages = {8},
          doi = {10.3847/1538-4357/ad47c1},
archivePrefix = {arXiv},
       eprint = {2311.07528},
 primaryClass = {astro-ph.SR},
       adsurl = {https://ui.adsabs.harvard.edu/abs/2024ApJ...969....8S}
}

@INPROCEEDINGS{Mohamed2010,
       author = {{Mohamed}, S. and {Podsiadlowski}, Ph.},
        title = "{Understanding Mass Transfer in Wind-Interacting Binaries: SPH Models of ``Wind Roche-lobe Overflow''}",
    booktitle = {International Conference on Binaries: in celebration of Ron Webbink's 65th Birthday},
         year = 2010,
       editor = {{Kalogera}, Vicky and {van der Sluys}, Marc},
       series = {American Institute of Physics Conference Series},
       volume = {1314},
        month = dec,
    publisher = {AIP},
        pages = {51-52},
          doi = {10.1063/1.3536409},
       adsurl = {https://ui.adsabs.harvard.edu/abs/2010AIPC.1314...51M}
}

@INPROCEEDINGS{Mohamed2007,
       author = {{Mohamed}, S. and {Podsiadlowski}, Ph.},
        title = "{Wind Roche-Lobe Overflow: a New Mass-Transfer Mode for Wide Binaries}",
    booktitle = {15th European Workshop on White Dwarfs},
         year = 2007,
       editor = {{Napiwotzki}, R. and {Burleigh}, M.~R.},
       series = {Astronomical Society of the Pacific Conference Series},
       volume = {372},
        month = sep,
        pages = {397},
       adsurl = {https://ui.adsabs.harvard.edu/abs/2007ASPC..372..397M}
}

@ARTICLE{Liub2024,
       author = {{Liu}, Boyuan and {Sartorio}, Nina S. and {Izzard}, Robert G. and {Fialkov}, Anastasia},
        title = "{Population synthesis of Be X-ray binaries: metallicity dependence of total X-ray outputs}",
      journal = {\mnras},
         year = 2024,
        month = jan,
       volume = {527},
       number = {3},
        pages = {5023-5048},
          doi = {10.1093/mnras/stad3475},
archivePrefix = {arXiv},
       eprint = {2308.06154},
 primaryClass = {astro-ph.HE},
       adsurl = {https://ui.adsabs.harvard.edu/abs/2024MNRAS.527.5023L}
}
\bibliographystyle{aasjournalv7}



\end{document}